\documentclass[prl,twocolumn,amsmath,amssymb,eqsecnum]{revtex4-1}

\usepackage[T1]{fontenc}
\usepackage[latin9]{inputenc}
\usepackage{color}
\usepackage{latexsym}
\usepackage{float}
\usepackage{graphicx}
\usepackage{times}   
\usepackage{esint}
\usepackage[unicode=true,pdfusetitle,
 bookmarks=false,colorlinks=true,citecolor=blue,urlcolor=blue,linkcolor=red]{hyperref}
\usepackage{dcolumn}
\usepackage{bm}
\usepackage{braket}
\usepackage{mathrsfs}
\usepackage{verbatim}
\usepackage{psfrag} 
\usepackage{amsmath,amssymb} 
\usepackage{colordvi} 
\usepackage{setspace}
\usepackage{grffile}
\usepackage{wasysym}
\usepackage{booktabs}
\usepackage{hyperref}
\usepackage{subfigure}
\usepackage{tabularx}
\setcitestyle{sort}
\usepackage[titletoc,toc]{appendix}

\newcommand\beq{\begin{equation} \begin{aligned}}
\newcommand\eeq{\end{aligned} \end{equation}}

\newcommand\bmb{\left( \begin{matrix}}
\newcommand\emb{\end{matrix} \right)}

\begin{document}

\title{Numerical evidence for a chiral spin liquid in the $XXZ$ antiferromagnetic Heisenberg 
model on the kagome lattice at $m=\frac{2}{3}$ magnetization}

\author{Krishna Kumar}
\author{Hitesh J. Changlani}
\author{Bryan K. Clark}
\author{Eduardo Fradkin}
\affiliation{Department of Physics and Institute for Condensed Matter Theory, 
University of Illinois at Urbana-Champaign, 1110 West Green Street, Urbana, Illinois 61801-3080}

\date{\today}

\begin{abstract}
We perform an exact diagonalization study of the spin-$1/2$ $XXZ$ Heisenberg 
antiferromagnet on the kagome lattice at finite magnetization $m = \frac{2}{3}$ with an emphasis on the XY point ($J_z = 0$), 
and in the presence of a small chiral term. Recent analytic work by Kumar, Sun and Fradkin [Phys. Rev. B 90, 174409 (2014)] 
on the same model, using a newly developed flux attachment transformation, predicts 
a plateau at this value of the  magnetization described by 
a chiral spin liquid (CSL) with a spin Hall conductance of $\sigma_{xy} = \frac{1}{2}$. 
Such a state is topological in nature, has a ground state degeneracy and 
exhibits fractional excitations. We analyze the degeneracy structure in the low energy manifold, identify the candidate 
topological states and use them to compute the modular matrices and Chern numbers all of which strongly agree with expected 
theoretical behavior for the $\sigma_{xy} = \frac{1}{2}$ CSL.  We argue that 
the evidence suggests the CSL is robust even in the limit of zero external chirality.
\end{abstract}
\maketitle

\section{Introduction}

Quantum spin liquids have been the focus of a great deal of research for over two decades and are expected 
to exist in frustrated quantum antiferromagnets. A natural model, both theoretically and experimentally, to look 
for these states is the spin-$\frac{1}{2}$ quantum Heisenberg antiferromagnet on the kagome lattice. 
The discovery of materials such as Herbertsmithite, which do not appear to order down to the lowest accessible 
temperatures~\cite{Helton2007,Mendels2007} and, at least qualitatively, are described by this simple model, 
has given new impulse to theoretical investigations~\cite{Yan2011,Depenbrock2012,Iqbal2011,Evenbly2010,Ran2007,Jiang2008,
Clark2013,Lauchli2011}. Relatively recent reviews on spin liquid phases in frustrated quantum antiferromagnets 
can be found in Refs.~\cite{Balents2010,Balents2016,Norman2016}.
\begin{figure}
\centering
\includegraphics[width=\linewidth]{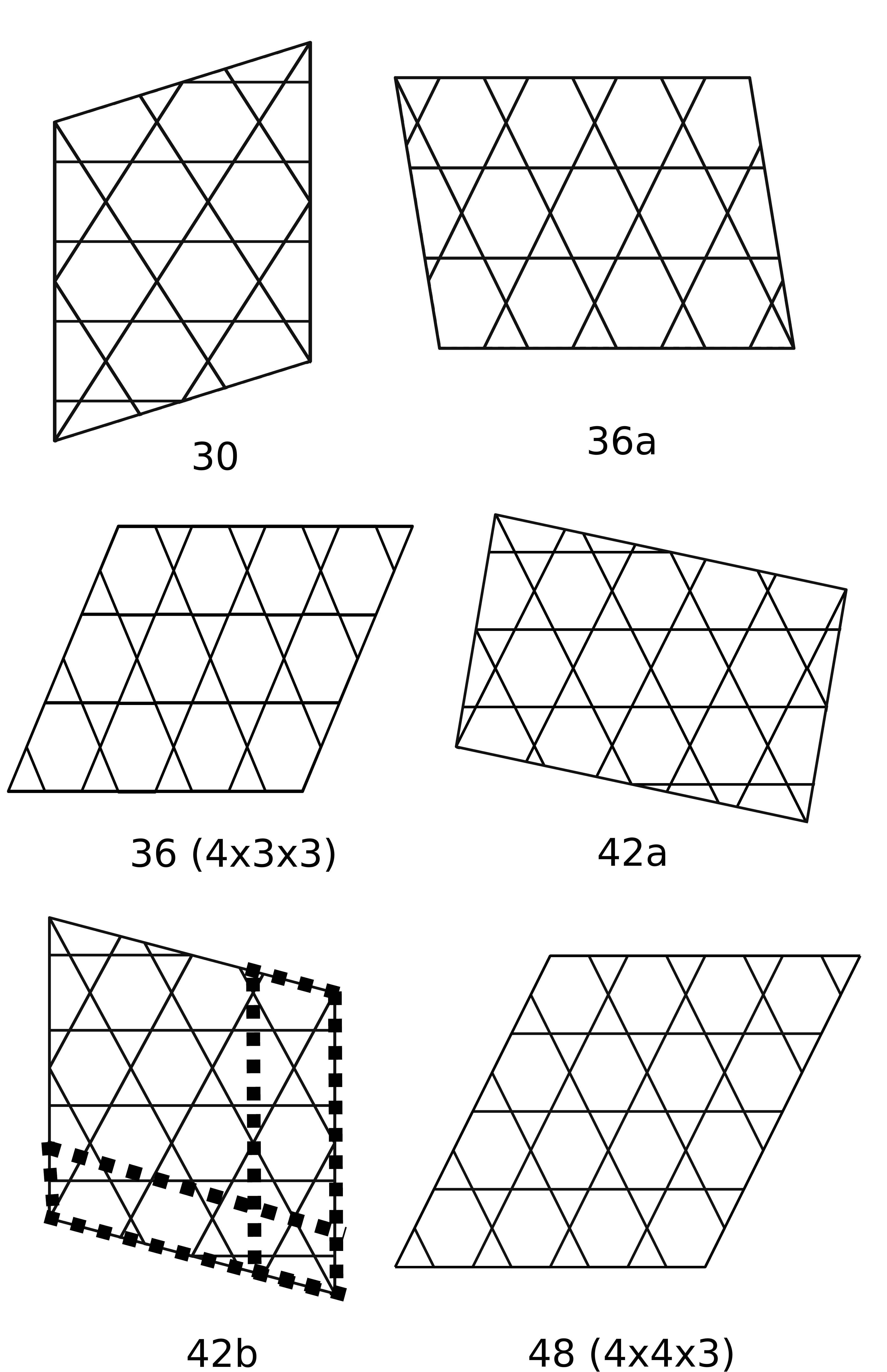}
\caption{Kagome clusters we study notated in the standard way (see Ref.~\cite{Lauchli2011}). 
The case of 36d, not shown here, 
is presented in Appendix~\ref{sec:analysis_36d}. Dotted black lines on $42$b site cluster 
correspond to two topologically non-trivial cuts used to compute the reduced density matrices and the corresponding MES states 
as outlined in Section~\ref{sec:mes}. Other clusters have similar cuts.} 
\label{fig:clusters}
\end{figure}

Several types of spin liquid states have been proposed to exist in quantum antiferromagnets~\cite{Zaletel2015} 
(of which the kagome lattice is a notable example), for example, topological states such as the time-reversal invariant 
$\mathbb{Z}_2$ spin liquid~\cite{Sachdev1991,Fradkin1979,Moessner2001b} (also known as the Toric Code state~\cite{Kitaev2003}) 
and the double-semion spin liquid~\cite{Freedman2004}, Dirac spin-liquid~\cite{Ran2007}, 
as well as chiral spin liquid (CSL)~\cite{Wen1989}. In the absence of an external (Zeeman) magnetic 
field there is good evidence for a $\mathbb{Z}_2$ spin liquid being the ground state~\cite{Yan2011,Depenbrock2012}, 
although other energetically competitive phases have also been reported~\cite{Ran2007,Iqbal2011,Clark2013}. 
On the other hand, a CSL has been found on addition of a term that explicitly 
breaks the chiral symmetry~\cite{Bauer2014,Kumar2015}. There is also evidence for the CSL phase 
in a Heisenberg antiferromagnet with further neighbor Ising interactions~\cite{He2014,Gong2014}.

Most spin liquids have been discovered in zero external magnetic field, and here we explore the alternate 
possibility of finding one in finite field. We thus study the spin-$\frac{1}{2}$ 
quantum $XXZ$ Heisenberg antiferromagnet on the 
kagome lattice (henceforth referred to as KAF) 
at finite magnetization. The Hamiltonian of the KAF is, 
\begin{equation}
  H_{XXZ} = \sum_{\langle i,j \rangle} J_{xy} \left( S^x_i S^x_j + S^y_i S^y_j \right) + J_z S^z_i S^z_j -h_B \sum_i S_i^z 
\label{eq:XXZ_Hamiltonian}
\end{equation}
where $J_{xy}$ (set to 1 in this paper) and $J_z$ are the 
strengths of the transverse and Ising terms respectively, and $h_B$ is the external field. 
Indices $i,j$ are used for sites, $\langle i,j \rangle$ 
refers to nearest neighbor pairs and $S^{x,y,z}$ refer to the usual spin-1/2 Pauli matrices.
Since the Hamiltonian conserves total $S_z$, it is also common to simply drop the $h_B$ term and instead work at 
fixed magnetization ($m$), defined as,
\begin{equation}
	m \equiv \frac{n_{\uparrow}-n_{\downarrow}} { n_{\uparrow} + n_{\downarrow} }
\end{equation}
where $n_{\uparrow}$ ($n_{\downarrow}$) are the number of up (down) spins. 

Several facets  of the model of Eq.\eqref{eq:XXZ_Hamiltonian} 
have been studied with different analytic and numerical techniques. For example, in the regime of strong 
Ising anisotropy, $J_z > J_{xy}$, a strong coupling expansion and exact diagonalization (ED) study has 
found evidence for valence-bond ordered phases at $m=1/3$ magnetization~\cite{Cabra2005}. 
At the Heisenberg point, $J_z = J_{xy}$, too ED~\cite{Capponi2013} 
and density-matrix renormalization group calculations~\cite{Nishimoto2013} 
have found strong evidence for magnetization plateaus  t $\frac{1}{3}, \frac{5}{9}$ and $\frac{7}{9}$ of the saturation value, 
and have further substantiated the existence of ordered phases. 

While the Heisenberg and Ising regimes appear to have been systematically explored, 
less is known about the $XY$ regime, i.e. close to $J_z = 0$. 
Analytic studies on the KAF in the $XY$ regime using 
flux attachment methods predict magnetization plateaus at $\frac{1}{3}$ and $\frac{5}{9}$ magnetization 
(just like the Heisenberg case) in addition to one at $\frac{2}{3}$ magnetization, 
all of which correspond to CSL states~\cite{Kumar2014}. These CSLs belong to 
the universality class of the Laughlin state for bosons with spin Hall conductance $\sigma_{xy}=\frac{1}{2}$ 
for the $m = \frac{1}{3}$ and $ m = \frac{2}{3}$ plateaus, and the first Jain state for bosons with 
$\sigma_{xy}=\frac{2}{3}$ for the plateau at $ m = \frac{5}{9}$~\cite{Kumar2014}. 
The CSL was postulated originally by Kalmeyer and Laughlin~\cite{Kalmeyer1987} 
as the ground state for the quantum antiferromagnetic spin-$\frac{1}{2}$ Heisenberg model on the 
triangular lattice. Note that the filling fraction of the corresponding fractional quantum Hall 
state is not equal to the filling fraction of the hard-core bosons (i.e. the magnetization fraction). 

In this paper, we focus our attention to the case of the state with magnetization $m=2/3$. 
This state is expected to be topological and described by an effective field theory 
with the form of a Chern-Simons gauge theory with gauge group $U(1)_k$ with level $k=2$, which is 
the effective field theory of a fractional quantum Hall effect for bosons at filling fraction $\nu=\frac{1}{2}$. 
(For a detailed review see Refs.~\cite{Wen1990,Wen1995,Fradkin2013}). 
The CSL has the following properties:  
(a) it does not break any symmetries aside from time reversal and parity, 
(b) it has a two-fold degenerate ground state on a 
torus (i.e. a system with periodic boundary conditions), 
(c) its elementary excitations are anyons with 
fractional statistics $\theta=\frac{\pi}{2}$ (hence, they are semions) and fractional ``charge'' $\frac{1}{2}$, 
and 
(d) on a system with an open boundary it has a chiral edge state described by a chiral compactified $U(1)$ 
boson conformal field theory (CFT), also at level $2$.
 
The topological properties of the CSL state on a torus are encoded in the modular $\mathcal{S}$ and $\mathcal{U}$ 
matrices which represent the response of the topological state to modular transformations 
of the torus~\cite{Wen1990}. In a general topological state, the matrix elements of the 
modular $\mathcal{S}$-matrix contain the quantum dimensions 
of the quasiparticles as well as their braiding properties, whereas the modular $\mathcal{U}$ matrix carries the information 
on the fractional spin, given in terms of the central charge of the associated chiral CFT and the conformal weight 
of the quasiparticles \cite{Witten1989,Dong2008}. In the case of the CSL (or the Laughlin state for bosons) 
there are two linearly independent  states on the torus, represented by the identity state $1$ and the Laughlin quasiparticle 
$\psi_{qp}$. The modular $\mathcal{S}$ and $\mathcal{U}$ matrices are  given by the $2 \times 2$ matrices \cite{Difrancesco-1997}
 \begin{equation}
\mathcal{S}=\frac{1}{\sqrt{2}}
 \bmb
 1 & 1\\
 1 & -1
 \emb
 ,\qquad
 \mathcal{U}=e^{i \frac{2\pi}{24}1}
 \bmb
 1 & 0\\
 0 & i
 \emb
 \label{eq:modular}
 \end{equation}
The factor of $\frac{1}{\sqrt{2}}$ in the $\mathcal{S}$ matrix 
is related to the effective total quantum dimension $D = \sqrt{2}$. \cite{Kitaev2006} 
This also specifies the ground state degeneracy of such a state, which is two in this case. 
The phase of the $\mathcal{S}_{22}$ term can be related to the phase obtained by braiding two quasiparticles around one another. 
The value of $-1$ indicates that the effective quasiparticles are semions. The modular $\mathcal{U}$ matrix encodes the value of 
the central charge (which is $c=1$ for a field theory expected to describe this phase) and the diagonal entries of the matrix 
give the phases that the various particles in such a system pick up under an exchange (for semionic quasiparticles 
it is $i= e^{i\frac{\pi}{2}}$). Hence, the modular matrices characterize the type of theory 
and the properties of the effective excitations or quasiparticles in the state.
 
Here we provide numerical evidence that the CSL is indeed realized in the KAF at $m=2/3$. 
Several properties of the low-energy states
are analyzed and elaborated upon in later sections of the paper, obtained using 
exact diagonalization of the KAF Hamiltonian of Eq.\eqref{eq:XXZ_Hamiltonian}. 
In order to aid the analysis, a 
small chiral term is added to the KAF Hamiltonian which explicitly breaks the time-reversal symmetry, allowing 
us to probe a single chiral sector consisting of two topologically related states. We thus 
establish and verify analytical results that argued in favor of a CSL in 
the regime of XY anisotropy for $m = \frac{2}{3}$; amongst these is 
the result that the many-body Chern number and hence the 
spin Hall conductance is $\sigma_{xy}=\frac{1}{2}$~\cite{Kumar2014} and
that the modular  $\mathcal{S}$ and $\mathcal{U}$ matrices match those expected
of a CSL. 

The magnetization $m = \frac{2}{3}$ corresponds to a high total $S_z$ sector, 
thereby restricting the Hilbert space and making the ED analysis feasible 
for reasonably large system sizes. To this end, we diagonalize the Hamiltonian 
of the KAF on finite clusters (discussed further in Section~\ref{sec:XXZ}) 
with up to 48 spins, with periodic boundary conditions. 
In Section~\ref{sec:energy_spectra}, we discuss the energy spectra as a function of model parameters 
and identify the quasi-degenerate ground states showing that these states
are robust under flux pumping through the torus. 
Following this, in Section~\ref{sec:flux_pumping}, we compute 
the many-body Chern number of the topological manifold finding it to be 
consistently 1/2 per state.
In Section~\ref{sec:mes}, the topological states are used to determine the minimally entangled 
states along two different non-trivial topological cuts. 
This enables evaluation of the modular matrices which are then compared to the predictions from topological field theory. 
All findings corroborate the existence of the CSL.  We then consider in sec.~\ref{sec:BeyondZeroChi} 
and sec.~\ref{sec:BeyondXY} the effects on the CSL as $J_\chi \rightarrow 0$ and $J_z >0$ respectively. 
Finally, in Section~\ref{sec:conclusions} we conclude by summarizing our results. 

\section{Calculation Details}
\label{sec:XXZ}

As mentioned in the introduction, we focus our attention to the KAF Hamiltonian of Eq.\eqref{eq:XXZ_Hamiltonian} 
and implicitly choose $h_B$ so as to fix the magnetization to
$m=2/3$. This problem is equivalent to considering a system of 
interacting hard-core bosons on the kagome lattice at 1/6 filling. 
In this mapping of spins to hard-core bosons, the term corresponding to $J_{xy}$ is the kinetic energy and 
that corresponding to $J_z$ is a density-density interaction. 
Notice that for the {\em antiferromagnetic} sign of $J_{xy}$, the kinetic energy describes 
bosons with a $\pi$ flux on every triangle of the kagome lattice and, hence, it is frustrated.
We focus primarily at the $XY$ point, 
$J_z=0$,  where analytical work suggests the existence of a CSL~\cite{Kumar2014}.  

While, the KAF is our primary interest, we have found it useful to consider the effects of adding 
a small chiral perturbation with strength $J_{\chi}$. 
Thus the model we study is, 
\begin{equation}
\begin{aligned}
H_{total} =& H_{XXZ} + H_{chiral} \\
H_{chiral} =& J_{\chi} \sum_{\triangle} \vec{S}_i \cdot \left( \vec{S}_j \times \vec{S}_k \right)
\end{aligned}
\label{eq:chiral}
\end{equation}
The chiral term is expected to naturally arise from the fermionic Hubbard model in a magnetic field 
and thus its inclusion is directly relevant to realistic materials. It must be kept in mind that 
the case of finite $J_{\chi}$ is not (in general) necessarily adiabatically connected 
to $J_{\chi} = 0 $. For example, in the KAF at the Heisenberg point in zero field, 
a phase transition from a $\mathbb{Z}_2$ spin liquid to CSL occurs on the introduction 
of a fairly small chiral term~\citep{Bauer2014}.

Our motivation for working with finite $J_{\chi}$ is primarily to probe a 
single (rather than both) chiral sector, which allows us to cleanly separate the 
topological states from their chiral (time-reversed) partners. This is needed since although the Zeeman 
term breaks time reversal symmetry, it does not break parity (i.e. mirror symmetry). Thus, a CSL state 
must break parity (chirality) spontaneously. The weak chiral term of Eq.\eqref{eq:chiral} breaks the chiral symmetry explicitly.
This greatly simplifies several analyses, especially when the energy scale corresponding to 
time reversal symmetry breaking is large~(on the scale of the gap of the topological manifold to the rest of the spectrum)  
and all four states cannot be clearly identified. Further subtleties associated with this identification 
will be discussed in later sections.

We analyze the model of Eq.\eqref{eq:chiral} using exact diagonalization of clusters of 
various sizes (ranging from $30$ to $48$ sites) and shapes, as is shown in Fig.\ref{fig:clusters}. 
Unless otherwise noted all clusters have periodic boundary conditions. 
Since there are several competing energy scales on the kagome lattice, the 
general extrapolation of properties to the thermodynamic limit is far 
from straightforward and each cluster requires individual 
consideration for detection of its topological manifold. In addition to the clusters shown in 
Fig.~\ref{fig:clusters}, we have performed calculations on cluster $36$d 
which we found to be similar to the 48 cluster,
discussed further in Appendix~\ref{sec:analysis_36d}.
 
\section{Energy spectra and ground state topological degeneracy}
\label{sec:energy_spectra}

\begin{figure*}
\centering
\includegraphics[width=.32\linewidth]{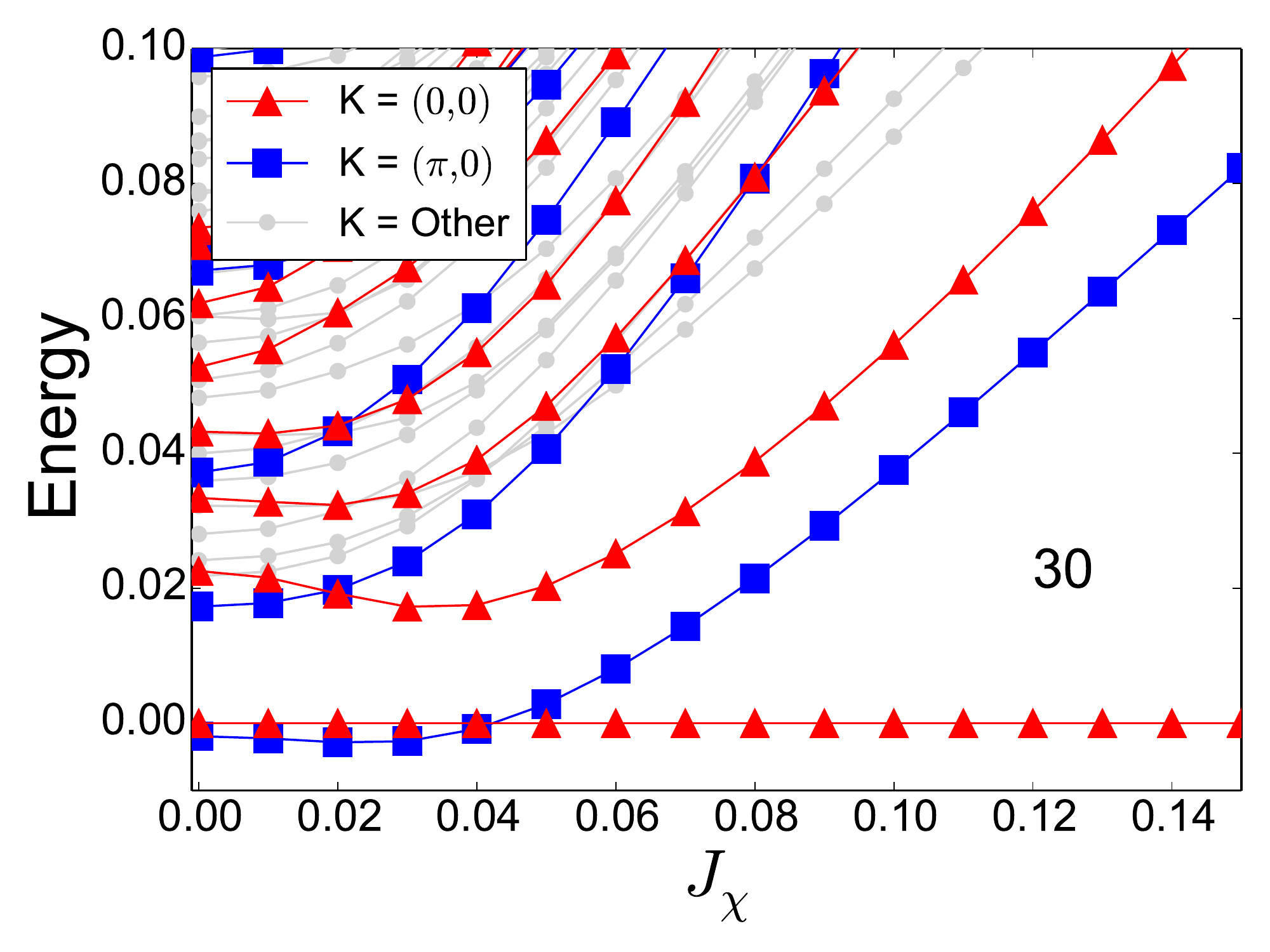}
\includegraphics[width=.32\linewidth]{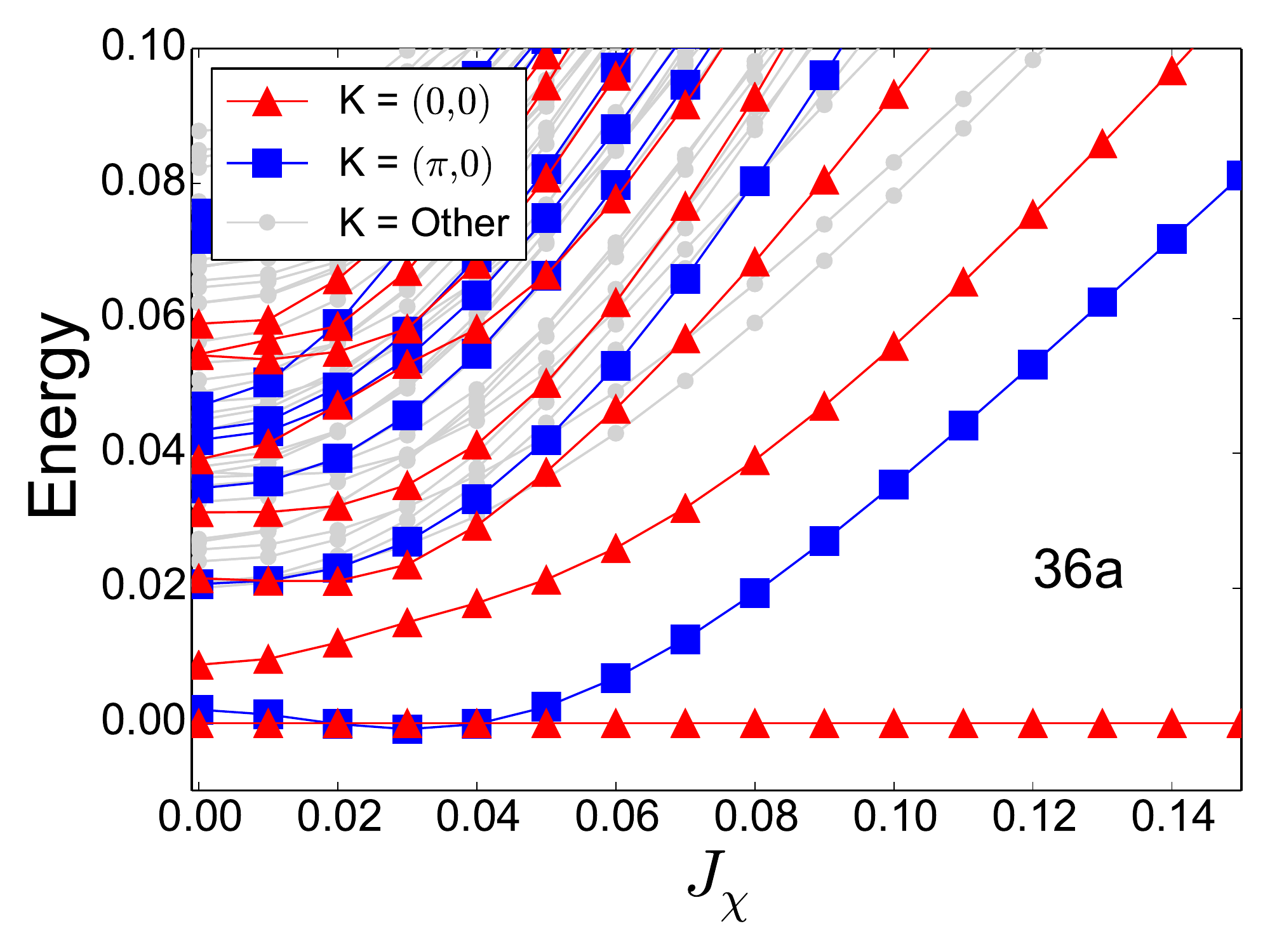}
\includegraphics[width=.32\linewidth]{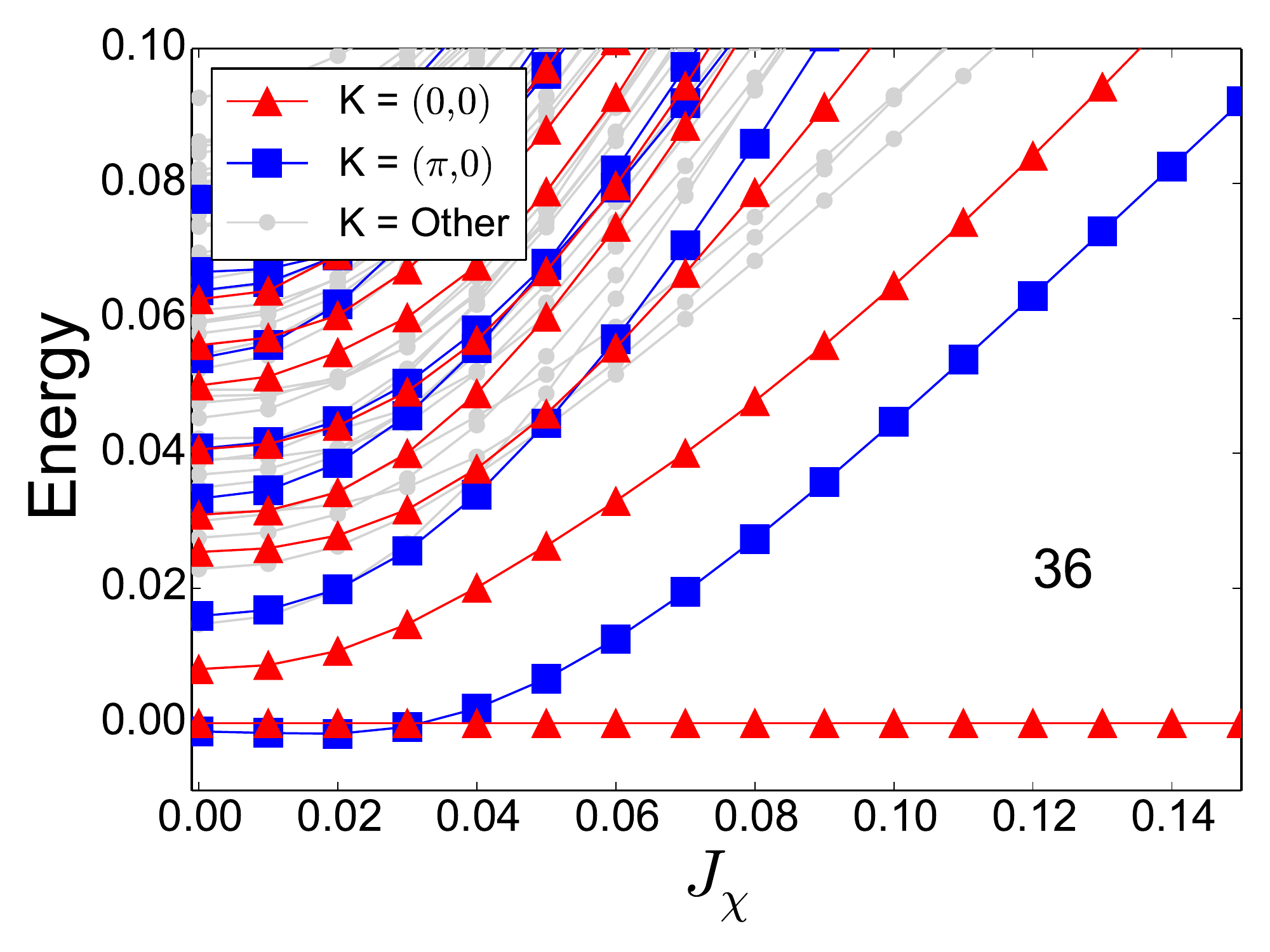}
\includegraphics[width=.32\linewidth]{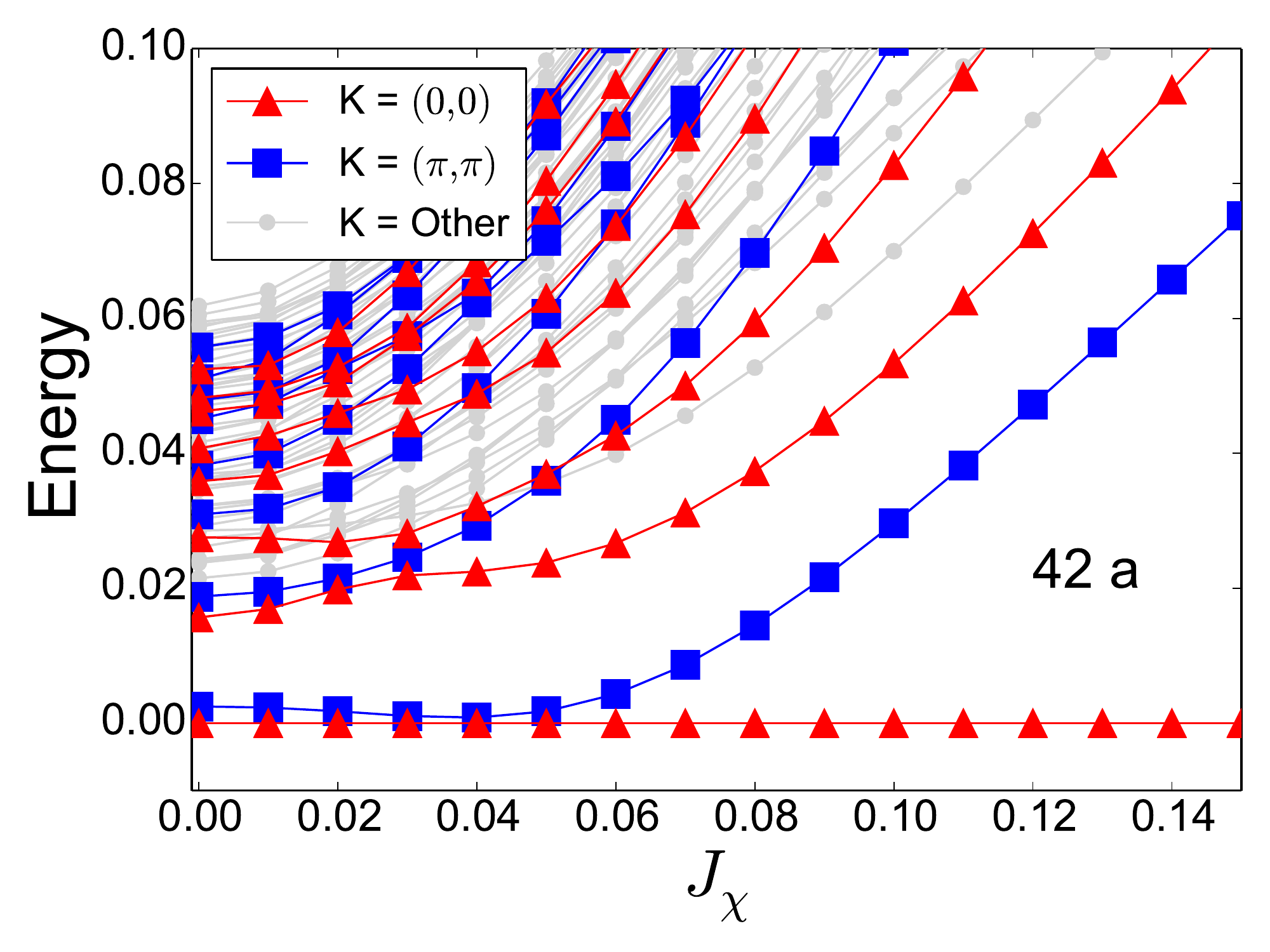}
\includegraphics[width=.32\linewidth]{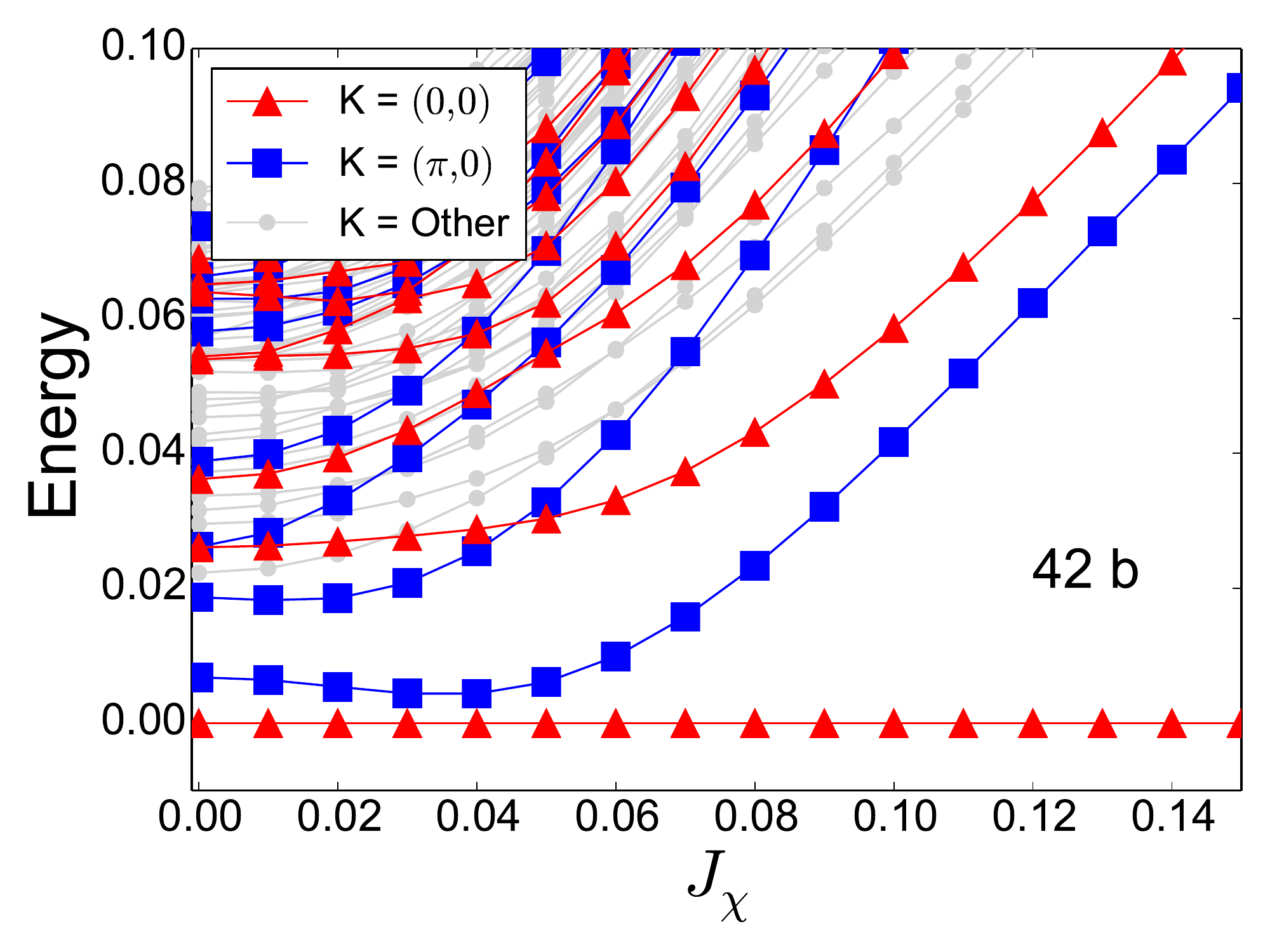}
\includegraphics[width=.32\linewidth]{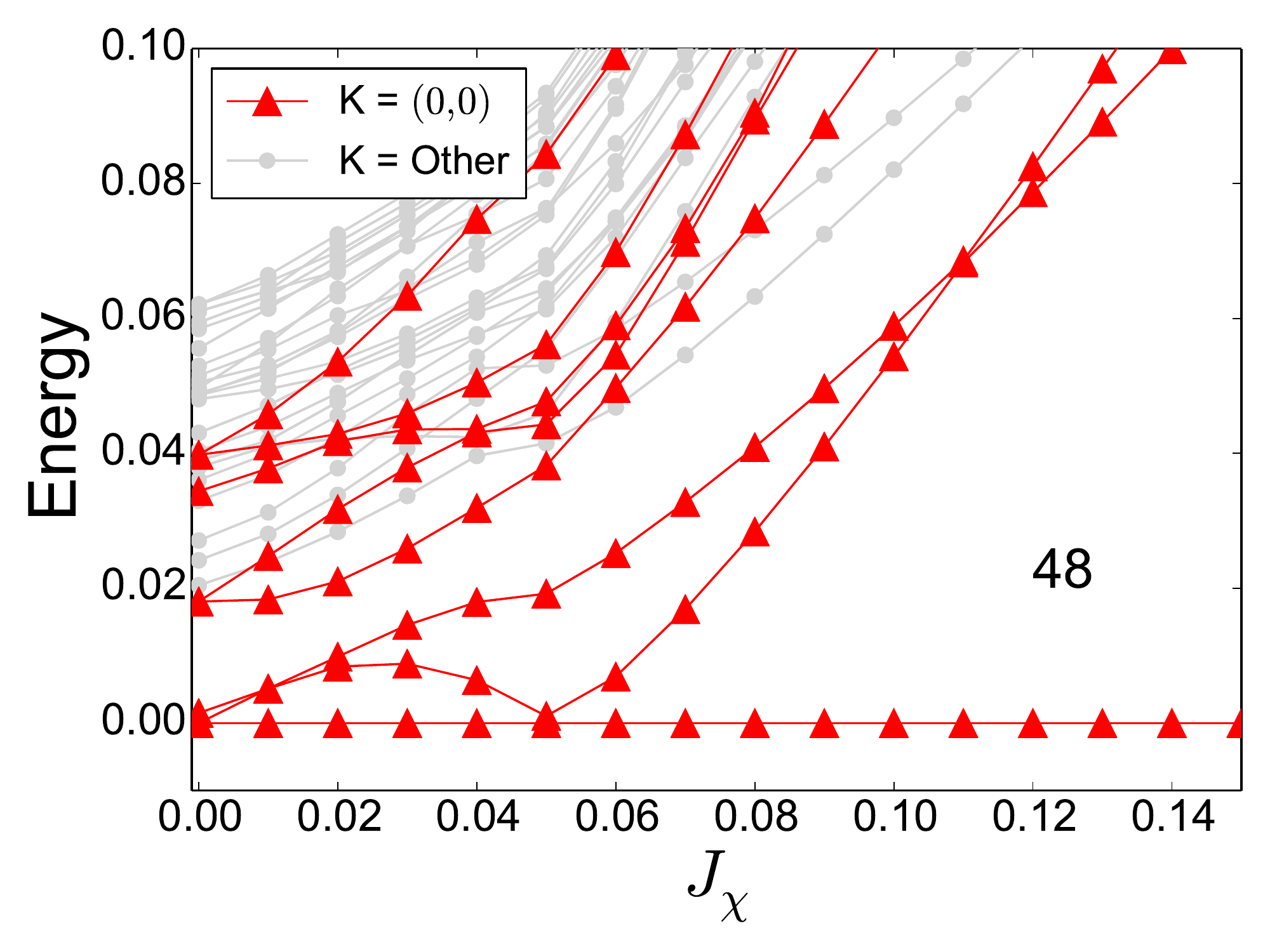}
\caption{Energy spectrum as a function of $J_{\chi}$ at the $XY$ point for clusters shown in Fig.~\ref{fig:clusters}.
At finite but small $J_{\chi}$ a two fold degeneracy expected from the existence of a CSL is seen. On approaching 
$J_{\chi} = 0$, a four fold degeneracy is expected, but owing to finite size effects no clear separation 
of this manifold from the rest of the states is seen on any of the clusters. 
}
\label{fig:cluster_energies_vs_Ch}
\end{figure*}

Chiral spin liquids on a torus have topological degeneracy in the thermodynamic limit. A 
CSL with $\sigma_{xy} = \frac{1}{2}$ has two topologically degenerate states per
chiral sector. By working at $J_\chi \approx 0.04$, we explicitly (but weakly) 
break the time-reversal symmetry of the real valued $XXZ$ Hamiltonian
and therefore would expect to see two nearly degenerate states if the system is a CSL. 
This is substantiated from our results in Fig.~\ref{fig:cluster_energies_vs_Ch} which shows 
the energy spectrum as a function of $J_{\chi}$ at the $XY$ point ($J_z = 0.0$) for various clusters. 
On all clusters, for small values of $0.02<J_{\chi}<0.05$, we observe a clear two-fold quasi-degeneracy 
in the energy spectrum which is well separated in energy from the next highest state
for all clusters presented. (See Table~\ref{table:cluster_summary} for the finite size energy gap 
to the closest and second-closest energy.) On most clusters, this feature persists even 
up to $J_{\chi}=0$. For $J_{\chi} > 0.07$, the quasi-degeneracy ceases to exist suggesting 
the occurrence of a quantum phase transition.  In addition, on the 48 site cluster, we see
signs of a phase transition as a discontinuity in the first derivative of the energy at 
$J_{\chi}=0.053$.

\begin{table*}
\setlength\extrarowheight{8pt}
\begin{ruledtabular}
 \begin{tabular}{cccccc}
  Clusters	&  Topological Gap & Excitation Gap &	Chiral expectation & $\mathcal{S}$-matrix  & Chern No. per state
  \\ \specialrule{1.1pt}{1pt}{1pt} 
  30		&	$0.000854$	& $ 0.018358$
  		&  $\begin{matrix} 	\langle \chi_{(0,0)} \rangle = 0.0684 \\
  						 	\langle \chi_{(\pi,0)} \rangle = 0.0545 \end{matrix}$	
  		&  $\bmb 0.697 & 0.701 \\ 0.717 & -0.713 \emb$
		& $ \begin{matrix} \frac{1}{2}   \\ 
				   \frac{1}{2}	 \end{matrix}$ \\
  $36$a	&	0.000142 & 0.017929
  		&  $\begin{matrix} 	\langle \chi_{(0,0)} \rangle = 0.0676 \\
  						 	\langle \chi_{(\pi,0)} \rangle = 0.0608 \end{matrix}$	
  		&  $\bmb 0.706 & 0.708 \\ 0.708 & -0.706 \emb$
		& $ \begin{matrix} \frac{1}{2}   \\ 
				   \frac{1}{2}	 \end{matrix}$ \\
  $36$		&	0.002186 & 0.020074
  		&  $\begin{matrix} 	\langle \chi_{(0,0)} \rangle = 0.0705 \\
  						 	\langle \chi_{(0,\pi)} \rangle = 0.0554 \end{matrix}$	
  		&  $\bmb 0.717 & 0.726 \\ 0.697 & -0.688 \emb$
		& $ \begin{matrix} \frac{1}{2}   \\ 
				   \frac{1}{2}	 \end{matrix}$ \\
  $36$d	&	0.004890 & 0.016572
  		&  $\begin{matrix} 	\langle \chi_{(0,0)_1} \rangle = 0.0558 \\
  						 	\langle \chi_{(0,0)_2} \rangle = 0.0537 \end{matrix}$	
  		& See Appendix~\ref{sec:analysis_36d} 
		&   1 ($\frac{1}{2}$ per state) \\
  $42$a	& 0.0008262	& 0.022473
  		&  $\begin{matrix} 	\langle \chi_{(0,0)} \rangle = 0.0631 \\
  						 	\langle \chi_{(\pi,\pi)} \rangle = 0.0622 \end{matrix}$	
  		& See Appendix~\ref{sec:42a_modular} 
		& $ \begin{matrix} 0.510   \\ 
				   0.490	 \end{matrix}$ \\
  $42$b	& 0.004374	& 0.025488
  		&  $\begin{matrix} 	\langle \chi_{(0,0)} \rangle = 0.0572 \\
  						 	\langle \chi_{(\pi,0)} \rangle = 0.0546 \end{matrix}$	
  		&  $\bmb 0.686 & 0.690 \\ 0.728 & -0.724 \emb$	
		& $ \begin{matrix} \frac{1}{2}   \\ 
				   \frac{1}{2}	 \end{matrix}$ \\
  48 	&	0.006372	& 0.017958
  		&  $\begin{matrix} 	\langle \chi_{(0,0)_{+}} \rangle = 0.0531 \\
  						 	\langle \chi_{(0,0)_{-}} \rangle = 0.0653 \end{matrix}$	
  		&  $\bmb 0.705 & 0.694 \\
				0.694 & -0.736 e^{-i 0.088} \emb$	
		&  1 ($\frac{1}{2}$ per state)

  \end{tabular}
 \end{ruledtabular}
 \caption{Summary of results obtained on various clusters for $J_z = 0$ and $J_{\chi} = 0.04$. 
The chiral expectation values are computed on an elementary triangle of the kagome lattice and the sub-labels in the expectation correspond 
to the momentum (and rotation when applicable) eigenvalues  of the candidate topological states. 
The modular matrices shown above are obtained using the same procedure described in the main text around Eq.~\eqref{eq:S_42b_chi} 
and \eqref{eq:S_48_chi}. }
 \label{table:cluster_summary}
\end{table*}

For clusters $30$, $36$ and $36$a and $42$b the two quasi-degenerate states are in momentum sectors $K = (0,0)$ 
and $K = (\pi, 0)$ (or $K = (0,\pi)$ ) and for $42$a they are in $K=(0,0)$ and $K = (\pi,\pi)$.
For the rotationally symmetric $48$-site cluster, the two low energy states
lie in the $K = (0,0)$ momentum sector, 
consistent with the expectation that all the topological 
states are in $K = (0,0)$ when $N_p/N_x$ and $N_p/N_y$ are integers, where $N_p$ is the number of particles and 
$N_x$ and $N_y$ are the number of unit cells in the $x$ and $y$ directions, respectively~\cite{Wang2011}.
In $36$d, presented in Appendix~\ref{sec:analysis_36d}, the two lowest states are also in $K=(0,0)$.

Topological states are expected to be locally indistinguishable.  At $J_{\chi}=0.04$, we compare the chiral expectation 
values around a triangle $\Delta \equiv \{i,j,k \}$, 
\begin{equation}
	\chi_{\Delta} = \langle  \vec{S}_i \cdot \left( \vec{S}_j \times \vec{S}_k \right)  \rangle 
\end{equation}
of each of the two lowest states, finding that they are similar on all clusters 
with a value of approximately $0.06$ (see Table~\ref{table:cluster_summary}).

To probe the robustness of the proposed topological sector, we pump magnetic flux through non-trivial loops in the torus. 
Numerically, this is accomplished by twisting the boundary condition resulting
in the Hamiltonian $H(\theta_1,\theta_2)$ where the phase of $S^+_i S^-_j$ terms are modified so 
as to preserve the translational symmetry of the lattice and ensure that any non-trivial loop around direction $a$ 
of the lattice picks up total flux of $\theta_a$.

\begin{figure*}
\centering
\includegraphics[width=0.48\linewidth]{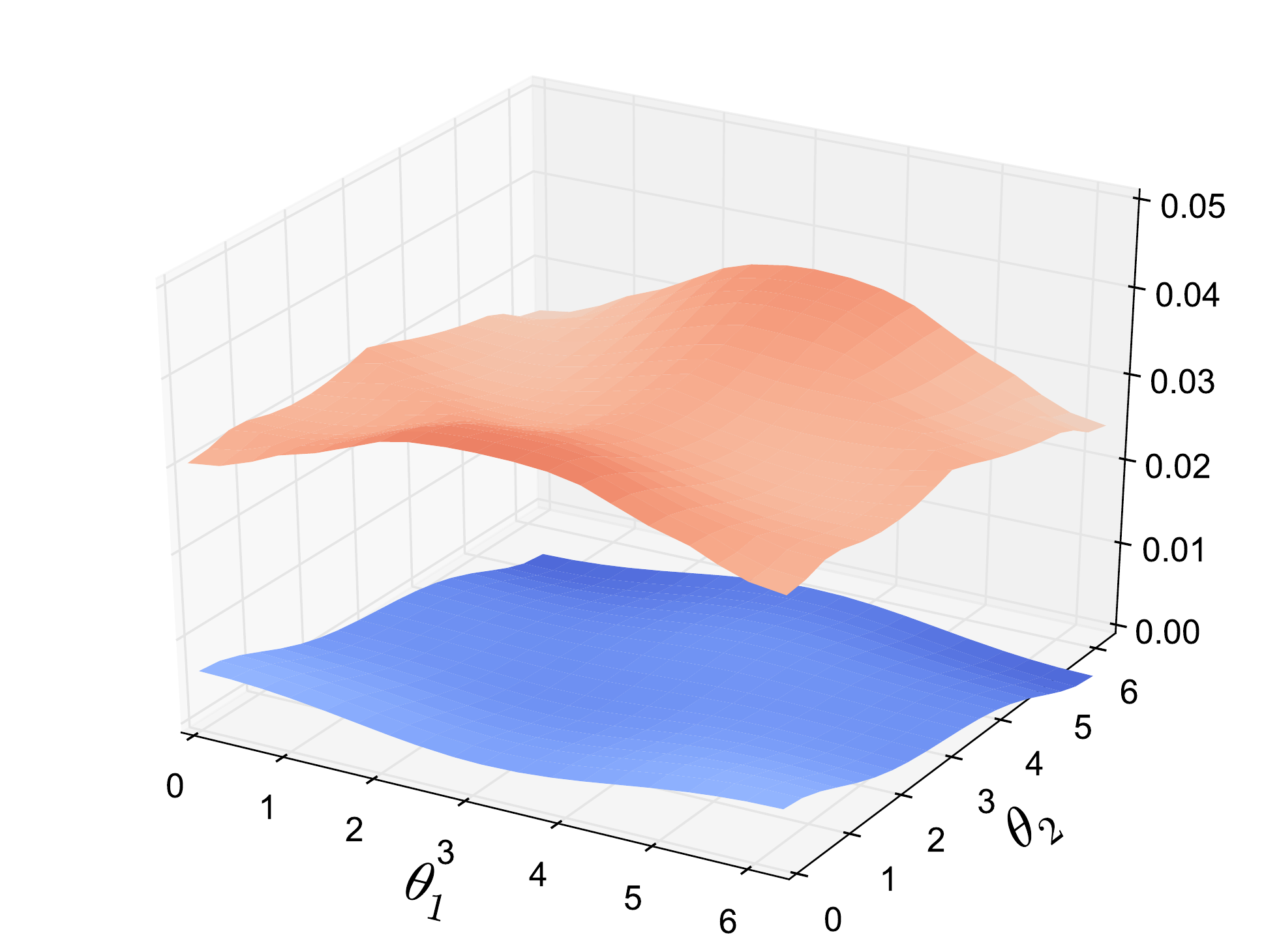}
\includegraphics[width=0.48\linewidth]{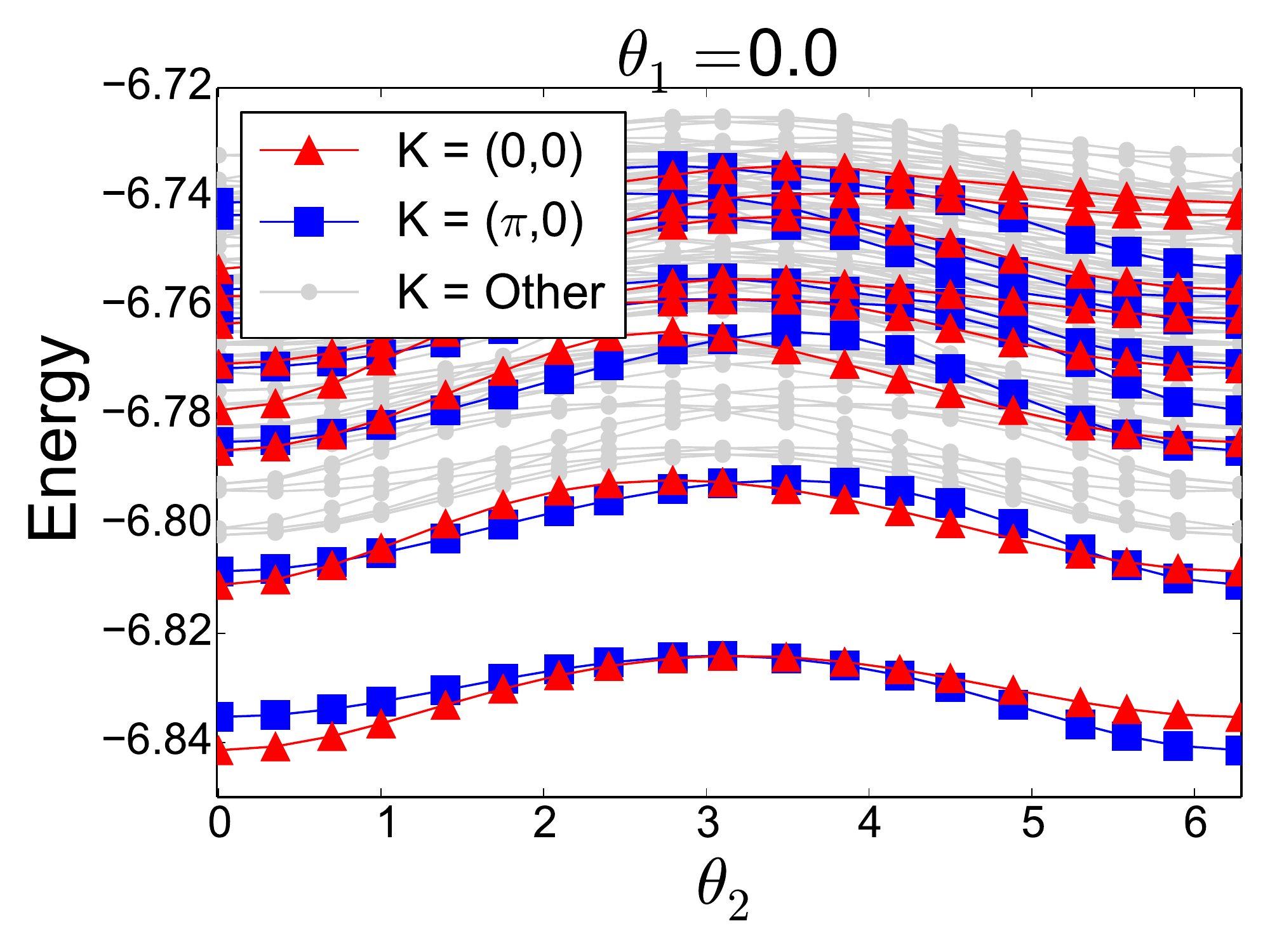}
\includegraphics[width=0.48\linewidth]{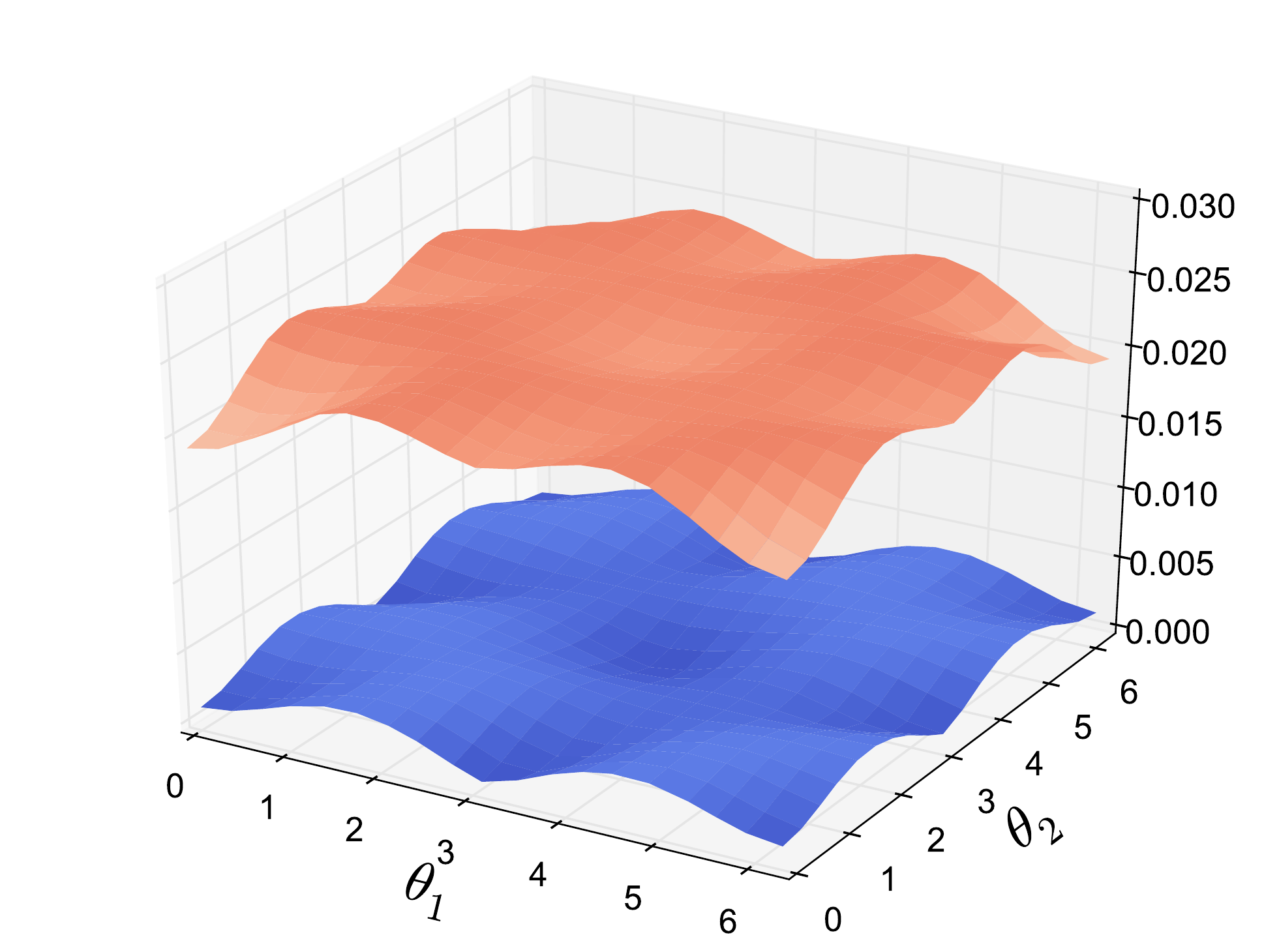}
\includegraphics[width=0.48\linewidth]{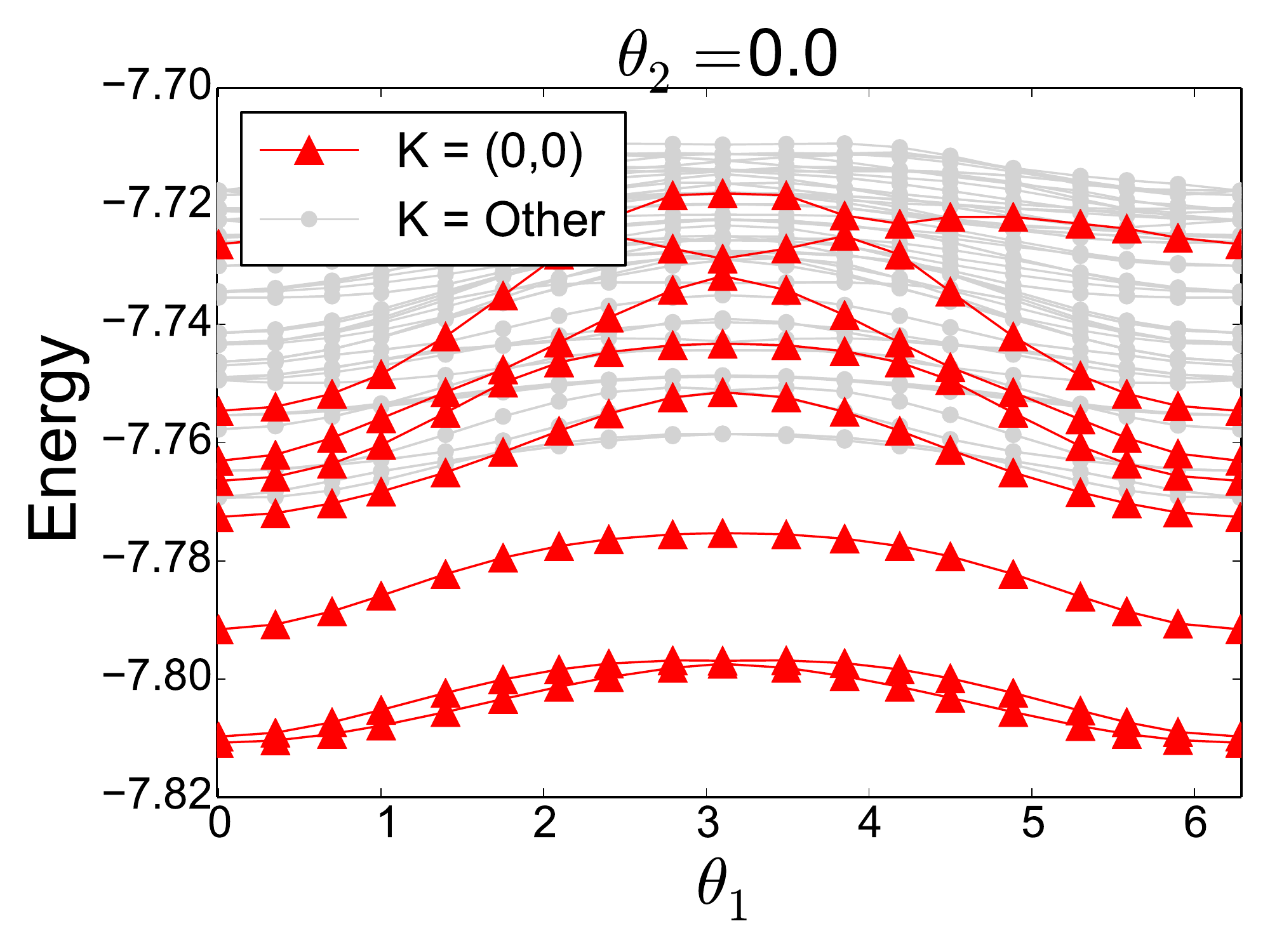}
\caption{Top: Energy difference of the $42$b cluster for $J_{\chi}=0.05$ and $J_z = 0 $ 
of the first three states with respect to the state at $K=(0,0)$ along  the two twist directions $(\theta_1,\theta_2)$ (left) and a representative cut of the energy of many states along $\theta_2$ at $\theta_1=0.0$ (right). 
The lowest $K = (0,0)$ state and $K = (\pi,0)$ states flip under flux pumping which is characteristic 
for degenerate states in different topological sectors. 
{Bottom:} Energy difference of the $48$ cluster for $J_{\chi}=0.05$ and $J_z = 0 $ 
of the first three states with respect to the lowest energy state along  the two twist directions $(\theta_1,\theta_2)$ (left) and a representative cut of the energy of many states along $\theta_1$ at $\theta_2=0.0$ (right).  
}     
\label{fig:42b_48_energy_spectrum_twist}
\end{figure*}

As is shown in Fig.~\ref{fig:42b_48_energy_spectrum_twist}, 
we consider a $20\times20$ grid of $(\theta_1,\theta_2)$ for 
the 42b and 48 cluster at  $J_\chi=0.05$ and find that the same two states are
non-trivially gapped across all twists. This establishes the clear separation of the 
topological manifold from the rest of the spectrum throughout the entire 
parameter space of flux pumps.

Fig.~\ref{fig:42b_48_energy_spectrum_twist} also shows the spectral flow of the low energy states 
as a function of $\theta_2$ from $0 \to 2\pi$ at fixed $\theta_1=0$. For the 42b cluster, 
we observe that the $K = (0,0)$ and $K = (\pi,0)$ momentum states exactly 
flip under the pumping of a $2\pi$ flux. This is characteristic of topological ground states 
and we will see in Sec.~\ref{sec:mes} that the individual ground states 
are minimally entangled states. Pumping a flux of $2\pi$ through a non-trivial loop on the torus allows for 
a transition from one topological sector to the other and 
the number of such $2\pi$ fluxes required for the system to return to its 
original state gives an indication of the ground state degeneracy. In this case we find two states
consistent with a CSL.

Notice that in the case of the $48$ site cluster, 
there is a third state that is not part of the topological manifold 
which is nonetheless separated from the rest of the continuum of low energy states. 
However, since it interlaces this manifold for small $J_{\chi}$, 
it is suggestive of its possibly non-trivial role in the low energy physics. 

Now that we have identified a topological manifold, 
we analyze our low energy wavefunctions in the next section and 
present evidence for the CSL state for small values of $J_{\chi}$ by computing the
Chern number in the topological manifold and the field-theoretic modular matrices 
mentioned in the introduction.

\section{Chern number}
\label{sec:flux_pumping}

Given the separation of the ground state topological manifold from the continuum, it 
is meaningful to compute the combined Chern numbers of the two topological ground states. 
The Chern number, which is a topological invariant, is defined as the integral 
of the Berry curvature over the two dimensional zone of twist angles as,
\beq
C = \frac{1}{2\pi} \int_0^{2\pi}\int_0^{2\pi} B(\theta_1, \theta_2) d\theta_1 d\theta_2
\label{eq:Chern_number_berry_flux}
\eeq
where $B(\theta_1, \theta_2)$ corresponds to the Berry curvature at the specific 
twist angles $\theta_1$ and $\theta_2$, which is defined as, 
\begin{equation}
	B (\theta_1,\theta_2) \equiv \partial_1 A_2 - \partial_2 A_1
\end{equation}
where $A_i (\theta_1,\theta_2)$ is the connection defined to be 
$A_i \equiv \langle \psi (\theta_1,\theta_2)| \partial_i | \psi (\theta_1,\theta_2)\rangle $
Numerically, this calculation is carried out by creating a grid of points in the $(\theta_1, \theta_2)$ space 
where $\theta_1 \in (0, 2\pi)$ and $\theta_2 \in (0, 2\pi)$ (in practice, the largest 
grid chosen for such computations was $20 \times 20$). 
Then the eigenvector $\ket{\psi(\theta_1,\theta_2)}$ is computed, 
and the Berry curvature at each point is calculated as,
\beq
B(\theta_1, \theta_2) = \text{Log} \big\{& \braket{\psi(\theta_1, \theta_2)|\psi(\theta_1 +\delta\theta_1, \theta_2) } \\
\times &\braket{\psi(\theta_1 + \delta\theta_1, \theta_2)|\psi(\theta_1 +\delta\theta_1, \theta_2 + \delta\theta_2) } \\
\times &\braket{\psi(\theta_1 + \delta\theta_1, \theta_2 + \delta\theta_2)|\psi(\theta_1, \theta_2 + \delta\theta_2) } \\
\times &\braket{\psi(\theta_1, \theta_2 + \delta\theta_2)|\psi(\theta_1, \theta_2) } \big\}
\eeq
which is essentially computing the overlap between wave functions along a closed loop in the $(\theta_1, \theta_2)$ grid. 
The $\delta\theta_1$ and $\delta\theta_2$ in the above expression refer to the grid spacing along the $\theta_1$ and $\theta_2$. 
When, the topologically related wavefunctions are in different 
momentum sectors they can be easily tracked individually and the gauge invariant quantity, 
i.e. the sum of the individual Chern numbers of the states can be computed 
directly by employing the formula above. When the two states cannot be (easily) tracked individually, 
such as the case on the $48$ site cluster where both topological states are in the $K=(0,0)$ sector, 
a Non-Abelian formulation is used~\cite{Yu2011,Shapourian2015}, which yields $B(\theta_1,\theta_2)$ 
to be the log of a product of determinants. 
\begin{figure}
\centering
\includegraphics[width=\linewidth]{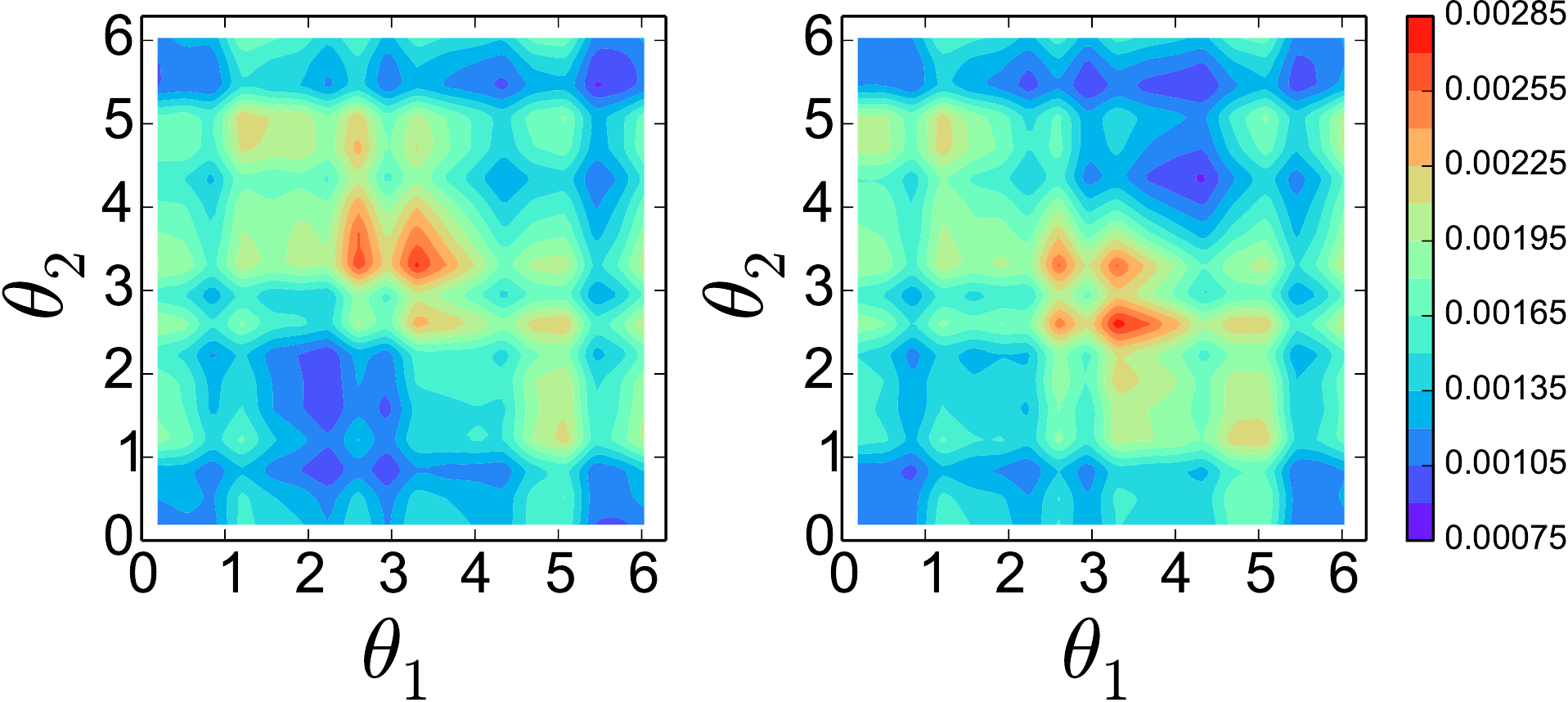}
\includegraphics[width=\linewidth]{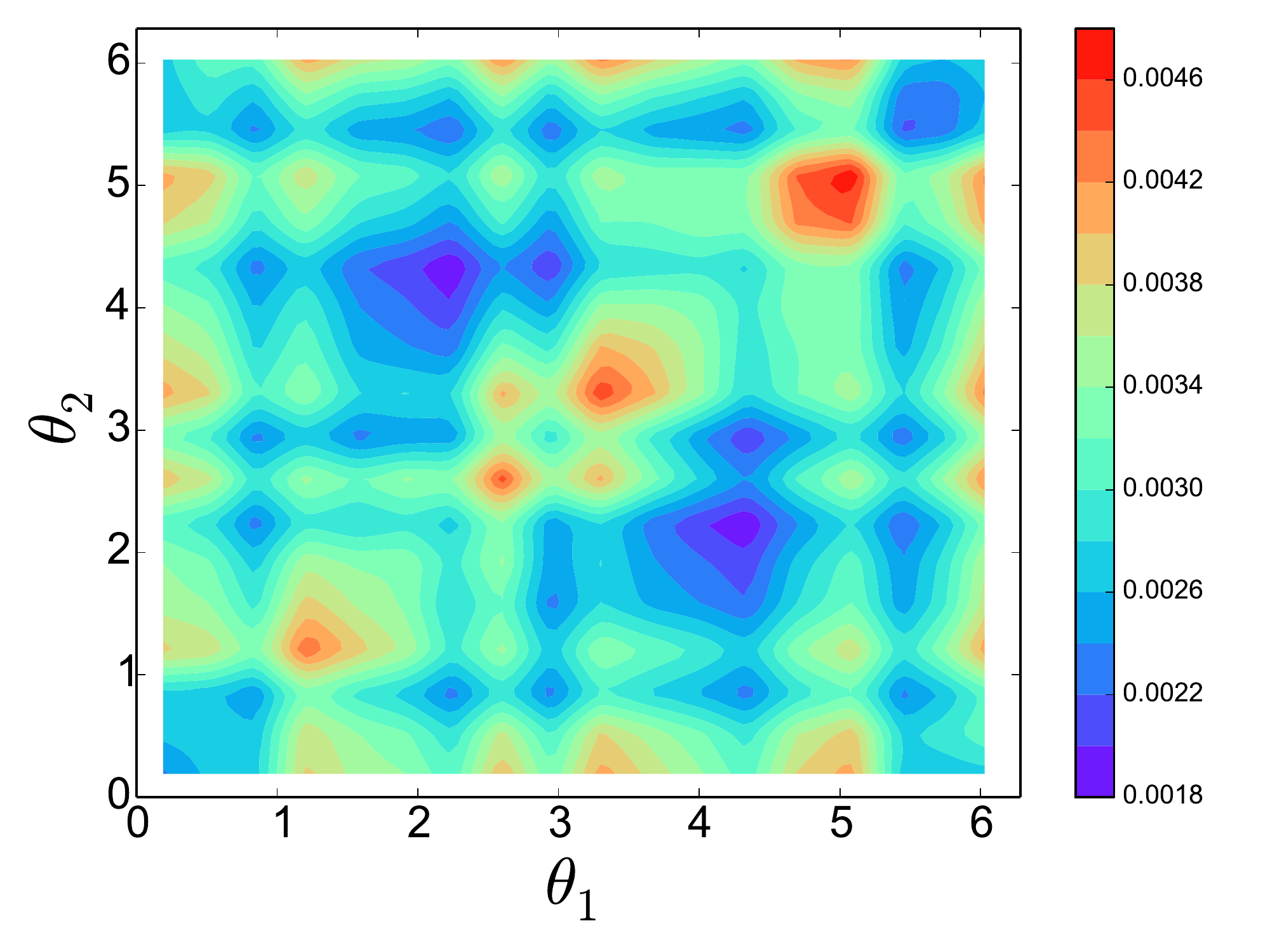}
\caption{
Distribution of Berry curvature for the $42$b $K = (0,0)$ (top left) 
and $K = (\pi,0)$ (top right) ground states and the two lowest $K=(0,0)$ 
states of the 48 site cluster (calculated together in a Non Abelian formalism), 
at $J_z=0$ and $J_{\chi}=0.05$. For all clusters, the Chern number 
sums to $+1$.}     
\label{fig:chern_distribution}
\end{figure}

The distribution of the Berry curvature, for $J_{\chi}=0.05$, 
is shown in Fig.~\ref{fig:chern_distribution} as a function of $\theta_1$ and $\theta_2$ 
for the $42$b and $48$ site clusters. For the 42b case, individual contributions 
from each topological state using Eq.~\eqref{eq:Chern_number_berry_flux} 
and for the $48$ site case the combined contribution of topological states from 
the Non-Abelian formalism, are shown.  The Chern number for the 
$K=(0,0)$ and $K=(\pi,0)$ states of the 42b cluster are 0.5 and 0.5 and 
the combined value of the two $K=(0,0)$ states of the 48 cluster is 1, at least to 6 decimal places. 
Chern numbers for all cases studied 
are summarized in Table~\ref{table:cluster_summary}.

We clarify that while the Chern number contributions from individual states can vary 
with $J_{\chi}$, their sum is a gauge invariant quantity which is constant 
as long as the states remain topological. For example, at $J_{\chi}=0.05$, 
the Chern numbers of the $K=(0,0)$ and $K=(\pi,\pi)$ states of the $42$a cluster 
are 0.507 and 0.493, and 0.55 and 0.45 at $J_{\chi} = 0.025$. However, 
in all cases, for small $J_{\chi}$, we find the same Chern number of $\frac{1}{2}$ per state. 
This is exactly the Chern number expected of a $\sigma_{xy} = \frac{1}{2}$ CSL. 

\section{Minimally entangled states and modular matrices}
\label{sec:mes}

In this section we focus on the $XY$ point in the presence of a small chiral term, 
specifically the results we present are for $J_{\chi} = 0.04$. 
On all clusters we use the lowest two quasi-degenerate states to compute the modular matrices 
following similar approaches used in Refs.~\cite{Zhang2012,Zhu2013, Cincio2012}. 
In this scheme, minimally entangled states (MES) are determined from the low-energy exact wavefunctions 
along topologically non-trivial cuts (see for example cluster 42b on Fig.\ref{fig:clusters}) 
and then appropriate wavefunction overlaps between these MESs are evaluated. 

We begin our analyses by considering the 
48 site cluster where all the low-energy states belong to the $K = (0,0)$ momentum sector.
The 48 site cluster is symmetric under a rotation of $\pi/3$ and thus 
we label the lowest two states ($\ket{\psi_{+}}$ and $\ket{\psi_{-}}$) 
by their rotational eigenvalues, found to be, 
$\lambda_+=\exp (i2\pi/3)$ and $\lambda_-=\exp (-i2\pi/3)$.
\begin{figure*}
\centering
\includegraphics[width=0.32\linewidth]{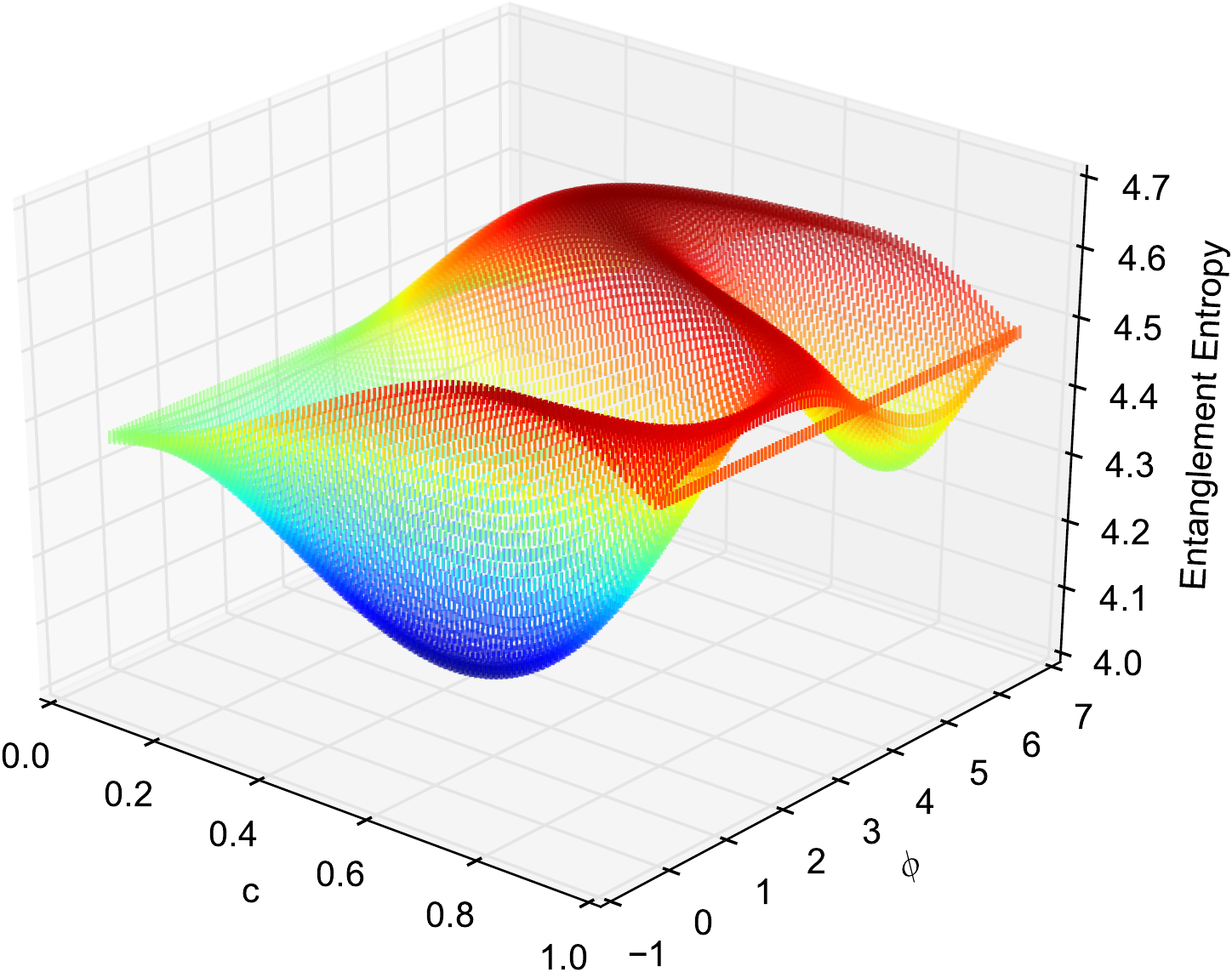}
\includegraphics[width=0.32\linewidth]{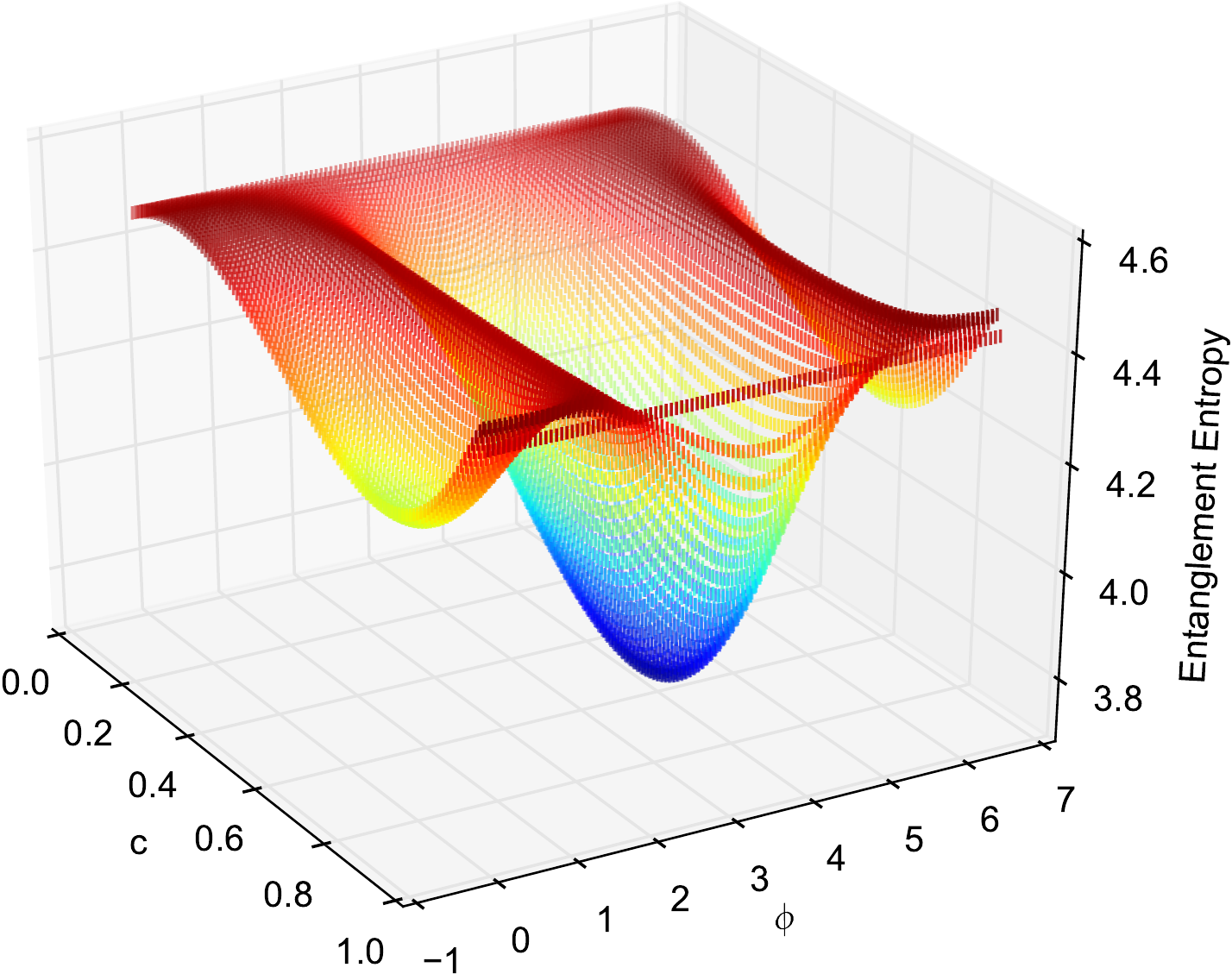}
\includegraphics[width=0.32\linewidth]{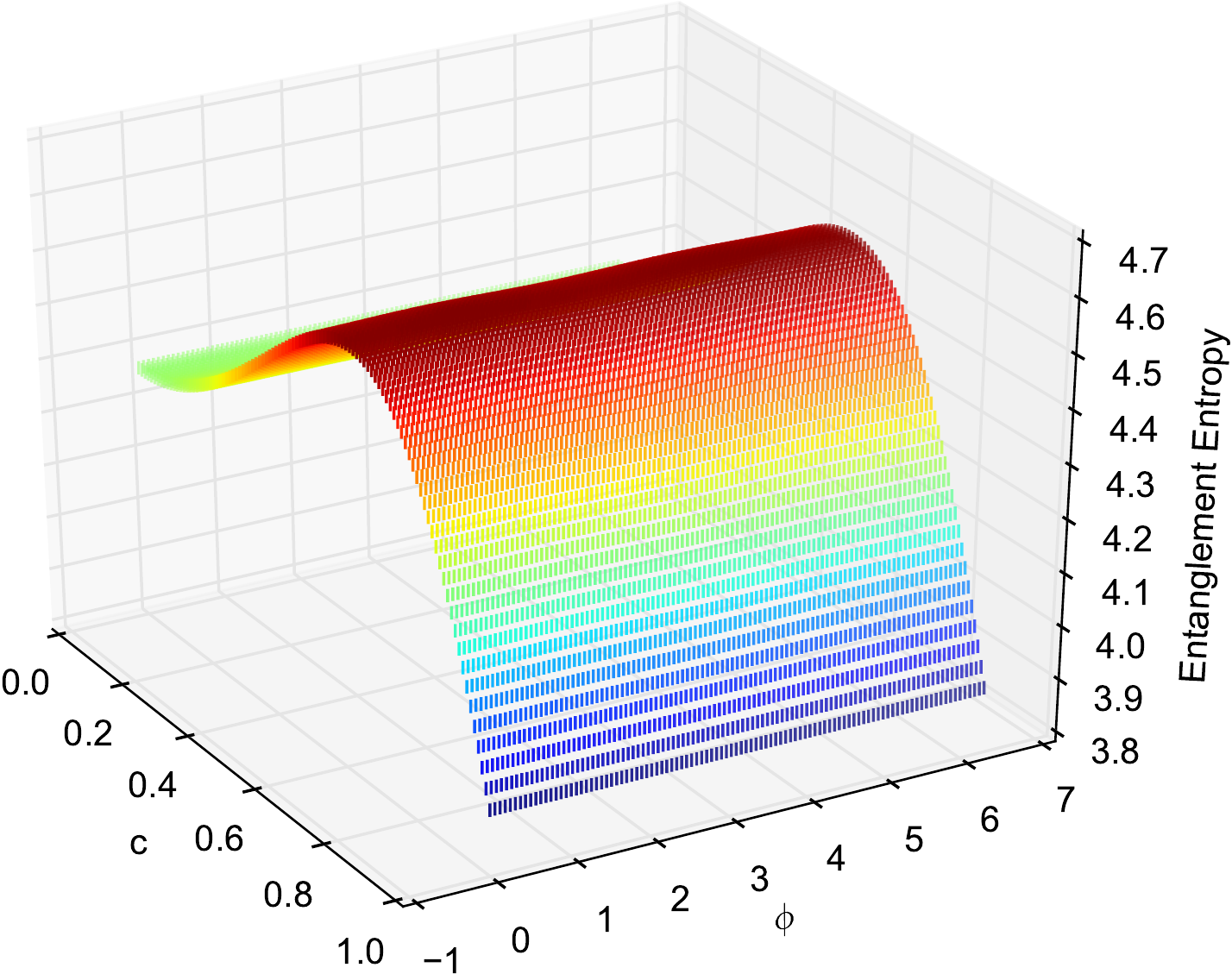}
\caption{Renyi entanglement entropy along topologically non trivial cuts as a function of 
$c,\phi$ for some representative cases. The entanglement for the 48 site cluster for 
$J_{\chi} = 0.04$ and $J_z = 0 $ (left) shows two minima 
(the other cut is related by rotational symmetry). For 
the 42b cluster (for $J_{\chi}=0$ and $J_z=0$) 
two minima are also seen for both cuts (center) 
along the vertical and (right) horizontal directions. 
(see Fig.~\ref{fig:clusters} for cut descriptions). Note that in the latter
case the minima coincide with eigenstates and occur at the boundaries of the 
two dimensional parameter space.}
\label{fig:42b_48_mes}
\end{figure*}
To compute the MES, we first construct the 
generic normalized linear combination of the two topological states,
\beq
\ket{\Psi} = c\ket{\psi_+} + \sqrt{1-c^2 } e^{i\phi} \ket{\psi_-}
\eeq
where $c \in [0,1]$ and $\phi \in (-\pi,\pi]$ are real parameters.
Next, we consider two non-trivial partitions (i.e. two distinct non-contractible Wilson loops) 
on the torus and obtain the second R\'enyi entanglement entropy $S_2$ 
from the reduced density matrices as follows
\begin{eqnarray}
	\rho_I & \equiv & \text{tr}_{I'} \ket{\Psi^+} \bra{\Psi^+} \\
	S_2    &\equiv  & -\ln  \text{tr} \rho_I^2 
	\label{eq:EE2}
\end{eqnarray}
where $\rho_I$ is the reduced density matrix for one of the cuts which we label as region $I$ and the trace $\text{tr}_{I'}$ 
runs over all the sites in region $I'$ ($I'$ corresponds to those sites that do not belong to the cut $I$). The 
MES states then correspond to local minima in the R\'enyi entanglement entropy in the $(c,\phi)$ parameter space.

For the 48 site cluster, the entanglement entropy as a function of $(c,\phi)$ is 
shown in Fig \ref{fig:42b_48_mes} for $J_z = 0.0$ and $J_{\chi} = 0.04$ along one non-trivial cut. 
We observe that there are indeed two local minima corresponding to,
\beq
\ket{\Psi^1} =& 0.462\ket{\psi_+} + 0.887 e^{i1.740} \ket{\psi_-} \\
\ket{\Psi^2} =& 0.901\ket{\psi_+} + 0.434 e^{i4.882} \ket{\psi_-}
\label{eq:48_mes_chi}
\eeq

The overlap between the above two states is $\braket{\Psi^1|\Psi^2} \approx 0.0315$, the small deviation 
from orthogonality being a finite size effect. Due the rotational symmetry of the 48 site cluster, 
one can obtain the MES states along the other cut by applying the rotation operator by $2\pi/3$ on the above MES states. 
This symmetry is special and allow us to compute both the modular 
$\mathcal{S}$ and $\mathcal{U}$ matrices, for which we follow the procedure outlined 
in other numerical works~\cite{Cincio2012} to carry out the calculation. Our result is,

\beq
\mathcal{S} =
\bmb
0.705 & 0.694 \\
0.694 & -0.736 e^{-i 0.088}
\emb
\label{eq:S_48_chi}
\eeq
and
\beq
\mathcal{U} = e^{i \frac{2\pi}{24} 1.014}
\bmb
1.000 & 0.000 \\
0.000 & i e^{0.053 i}
\emb
\eeq
The above modular matrices are in excellent agreement with the expected 
theoretical modular matrices shown in Eq.~\eqref{eq:modular}. 
The results for other clusters are similar; all results are 
summarized in Table~\ref{table:cluster_summary}.

We find similar modular matrices for  $0.02 \lesssim J_{\chi} \lesssim 0.07$ on the 48 site cluster
above which there appears to be a single ground state with a large gap to the excited states. 
Further, this gap tends to grow rapidly for larger values of $J_{\chi}$ as can be seen in Fig.~\ref{fig:cluster_energies_vs_Ch}. 

A similar behavior is also observed for other clusters 
(30, 36, and $42$b) for values of $J_{\chi} \lesssim 0.07$. On these clusters the topological ground states belong 
to different momentum sectors. Since these clusters lack rotational symmetry, we are restricted 
to only computing the modular $\mathcal{S}$ matrix. For example, for the $42$b site cluster when $J_{\chi} = 0.04$, 
the lowest two energy states are in momenta sectors $K=(0,0)$ and $K=(\pi,0)$.
respectively. Using these two topological states we get the below MES states along the two non-trivial cuts 
shown in Fig.~\ref{fig:42b_48_mes}. Along Cut I, we get
\beq
\ket{\Psi_{I,1}} =& 0.686 \ket{\psi_{(0,0)} } + 0.728 e^{i 2.639} \ket{\psi_{(\pi,0)}} \\
\ket{\Psi_{I,2}} =& 0.690 \ket{\psi_{(0,0)} } + 0.724 e^{i 5.781} \ket{\psi_{(\pi,0)}}
\label{eq:42b_chi_cutI}
\eeq
and the overlap between these two states along Cut $I$ is $\approx -0.0533$. Along Cut II we get 
\beq
\ket{\Psi_{II,1}} =& \ket{\psi_{(0,0)} } \\
\ket{\Psi_{II,2}} =& \ket{\psi_{(\pi,0)}}
\label{eq:42b_chi_cutII}
\eeq
which gives the below $\mathcal{S}$ matrix
\beq
\mathcal{S} = \bmb 0.686 & 0.690 \\ 0.728 & -0.724 \emb
\label{eq:S_42b_chi}
\eeq
 
The case of $42$a is similar to the above clusters in that the topological states 
exist in different momentum sectors. While the MESs along one cut are found 
to be in agreement with the theoretical expectation, the 
profile of the entanglement (as a function of $c,\phi$) along the other 
cut appears fairly flat, with only small variations in $S_2$ of the order of $0.1$. 
This makes the identification of the MES and hence the evaluation of the $\mathcal{S}$ matrix 
unreliable. This is attributed to a finite size effect of the relevant 
topologically non trivial loop and discussed further in 
Appendix~\ref{sec:42a_modular}. 
 
\section{Case of $J_{\chi} \rightarrow 0$}
\label{sec:BeyondZeroChi}
In the previous sections, we have presented strong numerical evidence for the existence of a CSL
phase in the $m=2/3$ sector with a small $J_\chi \approx 0.04$ chiral symmetry breaking term. 
In this section, we consider the limit where $J_{\chi} \rightarrow 0$. 

In this limit, our Hamiltonian is real and so (non-degenerate) states with non-zero chiral order must be spontaneously broken. 
This is true even though the introduction of a magnetic field has explicitly broken time-reversal symmetry in the spin language 
(there are more "up" than "down" spins); there is an additional time-reversal symmetry which is most evidently manifest in the bosonic
language where the symmetry corresponds to the operation of charge conjugation of the bosons.  Therefore, we anticipate 
four degenerate states in the thermodynamic limit: two topological states in each chiral sector.  
Unfortunately, as seen in Fig.~\ref{fig:cluster_energies_vs_Ch}, we do not observe four states
clearly separated from the continuum in any cluster. Therefore, either the finite-size breaking of the two chiral sectors
is large for these system sizes or the CSL does not survive in this limit. 

\textit{Most Clusters:} To explore this further, we first note that in all but the 48  and $36$d cluster (i.e. 30, 36a, 36, 42a, 42b) ,
the two nearly degenerate states stay gapped out from the rest of the spectrum as $J_{\chi}\rightarrow 0$ suggesting
the possibility that the finite-size breaking of the topological gap stays small while the chiral symmetry breaking gap is large.
As noted earlier, the chern numbers on the $42$a and $42$b clusters still are 1/2 per state at the lowest non-zero $J_\chi=0.025$ that
we explored. At $J_\chi=0$, on these clusters, we find that the all the contribution from the two-dimensional
twist space to the Berry phase is found to arise from a single "Berry monopole", a single point carrying all the Chern number. This
has a direct analog in the non-interacting band theory of topological insulators and 
is further discussed in Appendix~\ref{sec:dist_chern}.  The Chern number we compute in this situation is $\pm 1/2$ per state 
(the sign cannot be resolved because of the "Berry monopole."). 
Therefore, the Chern number for each state is still likely $1/2$ per state, although 
strictly speaking we cannot determine whether the gauge invariant sum of the
two topological states is zero or one.  

We compute the MES on the two lowest states, getting, in the 
case of the 42b cluster,
\beq
\mathcal{S} =
\bmb
0.700 & 0.74 \\
0.714 & -0.673
\emb
\label{eq:S_42b_Jz_0_Jchi_0}
\eeq

\begin{figure*}
\centering
\includegraphics[width=.32\linewidth]{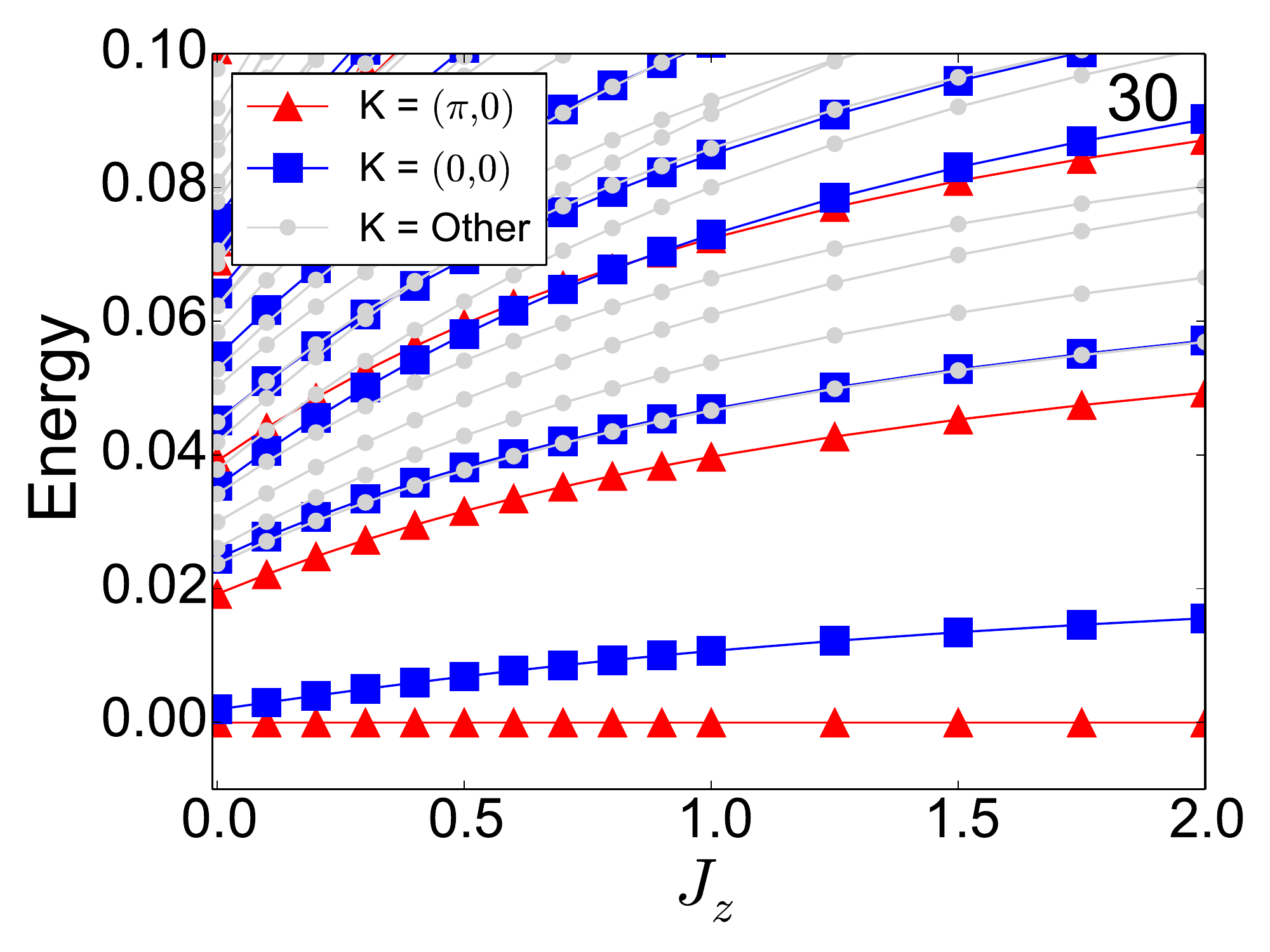}
\includegraphics[width=.32\linewidth]{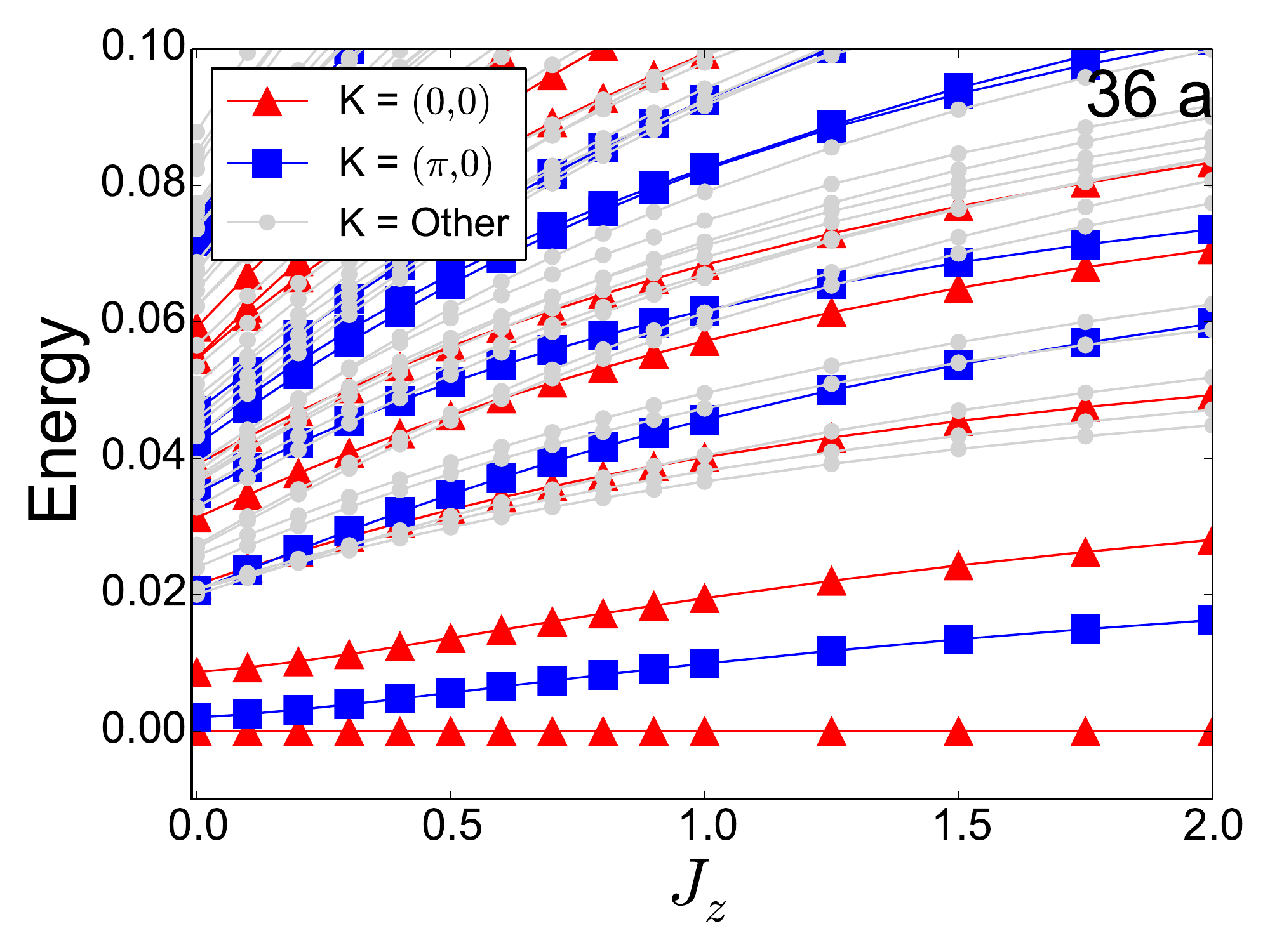}
\includegraphics[width=.32\linewidth]{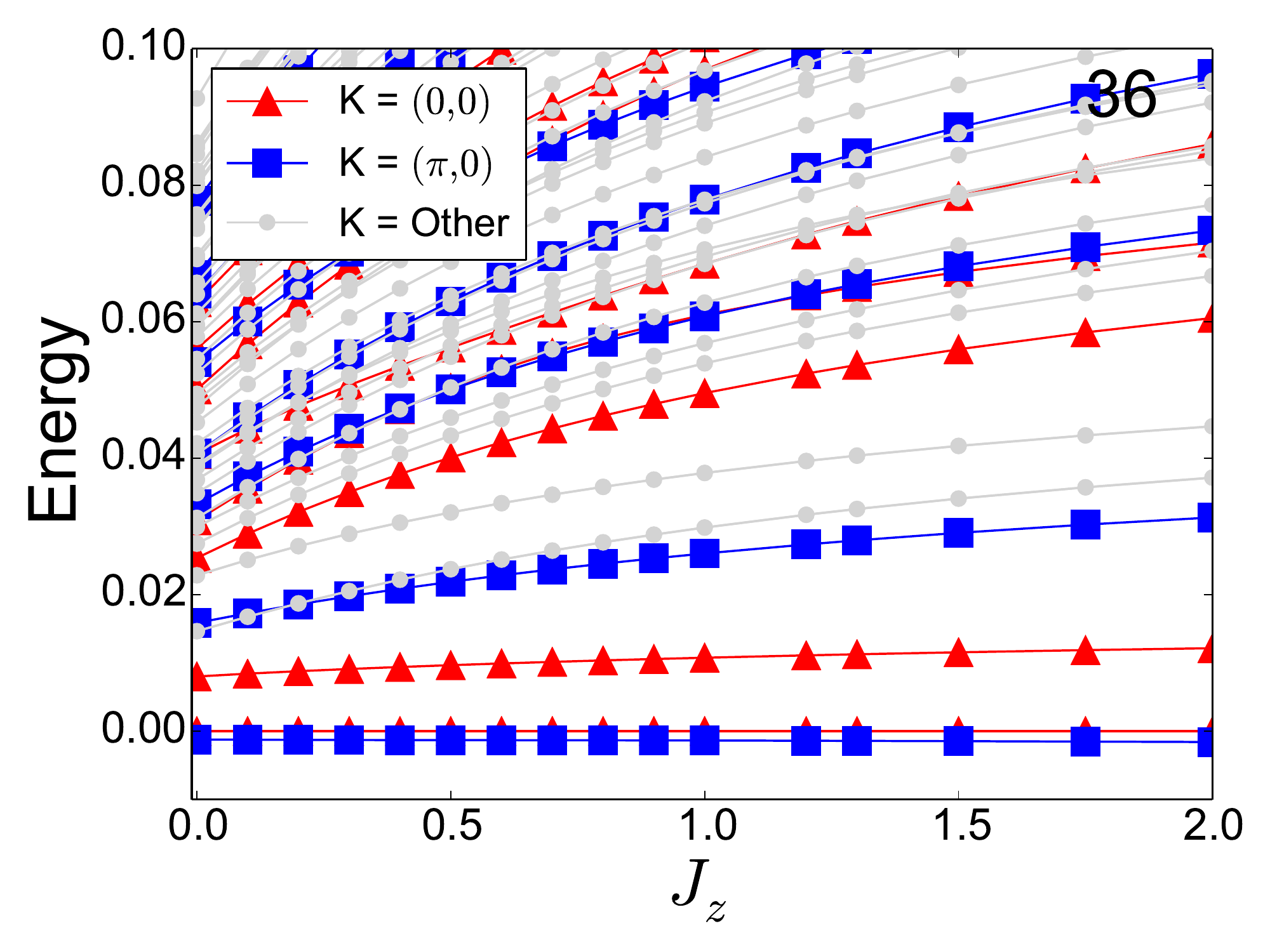}
\includegraphics[width=.32\linewidth]{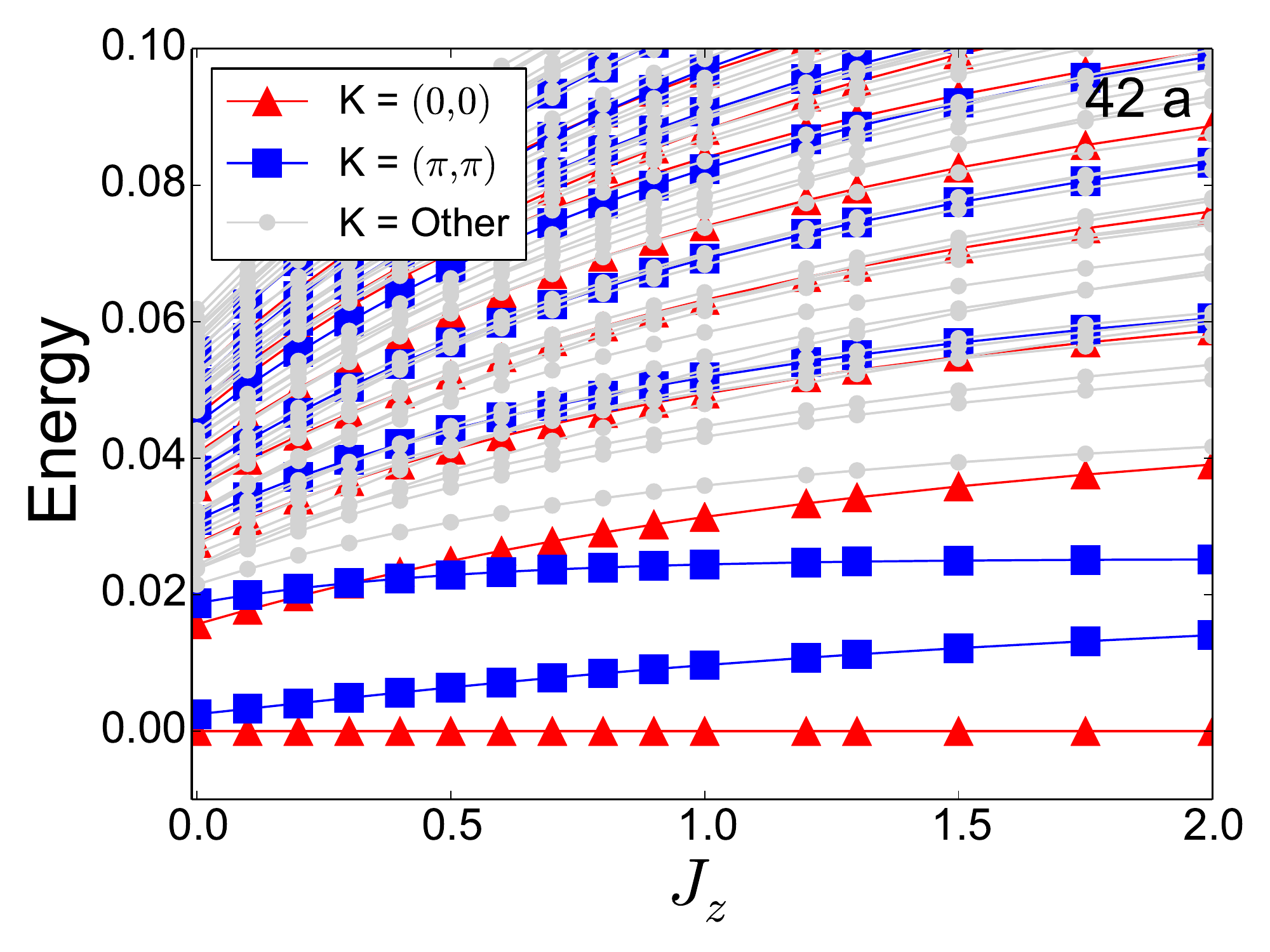}
\includegraphics[width=.32\linewidth]{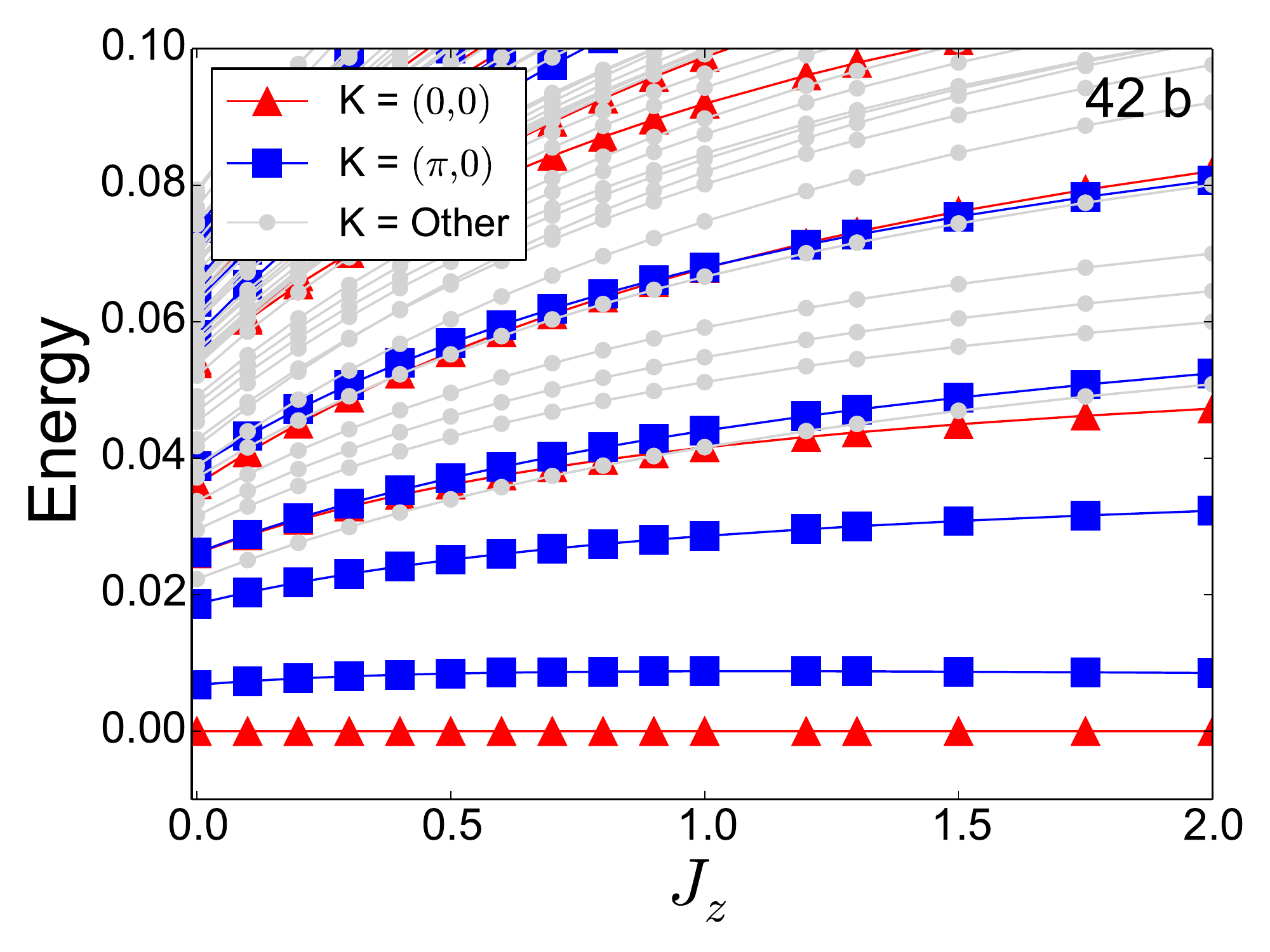}
\includegraphics[width=.32\linewidth]{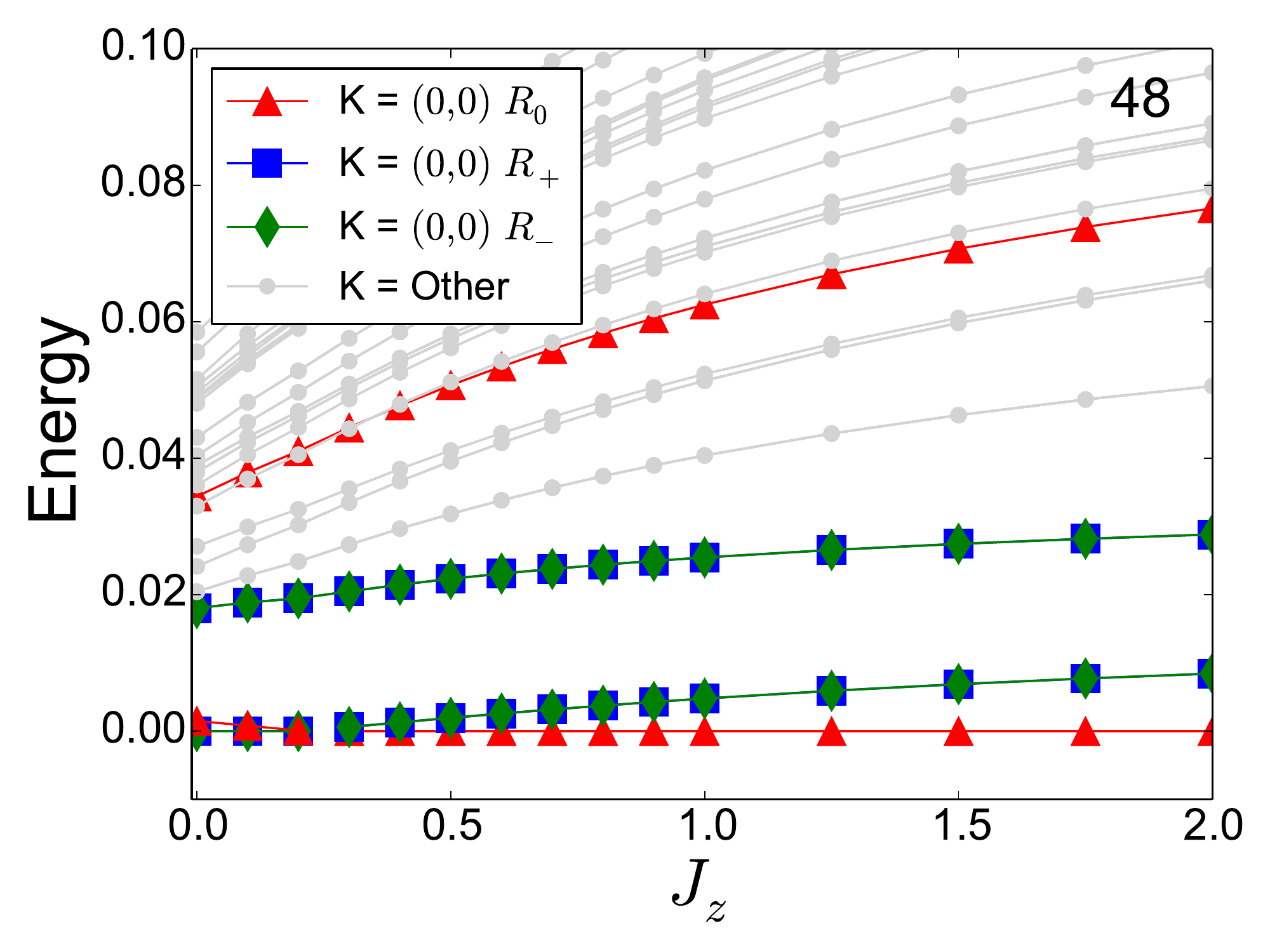}
\caption{Energy spectrum as a function of $J_{z}$ at $J_{\chi} = 0 $ for clusters shown 
in Fig.~\ref{fig:clusters}. For most clusters (except the $48$ site one), 
the low energy spectrum appears adiabatically connected to 
the large $J_z$ limit.}
\label{fig:cluster_energies_vs_Jz}
\end{figure*}

Once again, the above modular $\mathcal{S}$ matrix is in close agreement with the 
theoretically expected modular matrix consistent with the prediction that 
in the $XY$ limit the ground state at $m = \frac{2}{3}$ is a CSL with 
semionic quasiparticle excitations. 

We clarify that since these states are real and have no chiral expectation, 
the ideal scenario would be to identify the linear combination of 
states which form the chiral states and compute the MES on those. However, we find that these 
are located higher in the spectrum (and presumably mixed with other 
states), we do not attempt to make a definitive identification here. 

The fact that, on these clusters, the evidence for CSL seems to be robust down to $J_{\chi}=0$ 
suggests that the phase survives in this limit.  

\textit{The 48 cluster:}
The 48 cluster is more subtle as at $J_{\chi} = 0 $, we observe not only two exactly degenerate states but also a third state only 
slightly higher in energy. The two exactly degenerate ground states are labelled as $\ket{\psi_{\pm}}$ 
and the third state as $\ket{\psi_{0}}$ state due to their eigenvalues under the $\pi/3$ 
degree lattice-rotation operator i.e. $R_{\frac{\pi}{3}} \ket{\psi_{\pm} } = \lambda_{\pm}\ket{\psi_{\pm} }$ 
and $R_{\frac{\pi}{3}} \ket{\psi_{0} } = \lambda_0 \ket{\psi_{0} }$ 
where $\lambda_{\pm} = e^{\pm i \frac{2\pi}{3}}$ and $\lambda_0 = 1$. 
As expected of a real valued Hamiltonian, the two exactly ground degenerate states are complex conjugates of one another 
i.e. $\ket{\psi_{+}} = \ket{\psi_{-}}^*$; their chiralities are found to be 
approximately $+0.008$ ($\ket{\psi_{+}}$) and $-0.008$ ($\ket{\psi_{-}}$).

We find under twisting of the boundary conditions that the third nearly degenerate state does interlace with the 
other two low-energy states (among other states). Therefore we cannot cleanly identify a low-energy topological sector for which
we can compute Chern numbers or MES.  On this cluster, we therefore need a large enough 
value of $J_\chi$ before we can confidently compute the topological properties. 

\section{Away from the $XY$ point: $J_z > 0$}
\label{sec:BeyondXY}

While we have primarily focused on $J_z=0$, we will now consider the situation 
for $J_z > 0$ and $J_{\chi} = 0$. We show the spectrum as a function of $J_z$ in 
Fig.~\ref{fig:cluster_energies_vs_Jz}. For the non-48 site clusters, there is no indication 
of a gap closing at any value of $J_z$ out to $J_z \sim 2$ suggesting an adiabatic
connection to the Ising limit. For 42b we computed both the Chern number and MES 
throughout this entire range of $J_z$ and continue to find $C=1/2$ per state and two 
local MES whose $\mathcal{S}$ matrix matches the predicted CSL one
when taking the two lowest states.
The exception to this observation is the 48 site cluster 
where there is a clear energy crossing at $J_z \approx 0.2$ with the $\ket{ \psi_0 }$ becoming the ground state.
Given our current accessible system sizes, we are not able to arbitrate the true ground state phase at large $J_z$ 
given this disagreement.

\section{Conclusions}
\label{sec:conclusions}

In summary, using exact diagonalization, 
we have analyzed the low energy properties of the $XXZ$ antiferromagnetic Heisenberg 
model on the kagome lattice at finite magnetization $m=\frac{2}{3}$ magnetization 
with an emphasis on the $XY$ regime and in the presence of a small chiral perturbation. 
We were able to identify the topological states in the ground state manifold on multiple 
clusters and showed that they have the expected characteristics of the chiral spin liquid 
predicted by analytical calculations~\cite{Kumar2014}. We presented evidence of the chiral spin liquid with 
$\sigma_{xy} = \frac{1}{2}$ based on modular matrix computations which show the existence of semionic quasiparticles in these systems. 
We have also shown that flux pumping the states in different topological sectors evolve into one another as expected 
and finally also verified that the Chern numbers of the topological states are equal to $+1/2$. 
All of the above calculations provide strong evidence for the existence of the topological state 
in the presence of a small chirality term and that such a chiral state is very competitive with other states 
in the low energy spectrum even in the absence of the chirality term. 

\section{Acknowledgement}
\label{sec:ack}
We thank Hassan Shapourian and Victor Chua for illuminating discussions, 
especially regarding the subtleties of flux pumping and evaluation of the many-body Chern number.
HJC and BKC were supported by SciDAC-DOE grant DE-FG02-12ER46875.
This work was supported in part by the National Science Foundation grant No. DMR 1408713 (K.K. and E. F.) at the University of Illinois.
This research is part of the Blue Waters sustained petascale
computing project, which is supported by the
National Science Foundation (awards OCI-0725070 and
ACI-1238993) and the State of Illinois. Blue Waters
is a joint effort of the University of Illinois at Urbana Champaign
and its National Center for Supercomputing
Applications.

\begin{appendix}

\section{Modular matrices for the 42a case}
\label{sec:42a_modular}

\begin{figure}
\centering
\includegraphics[width=0.49\linewidth]{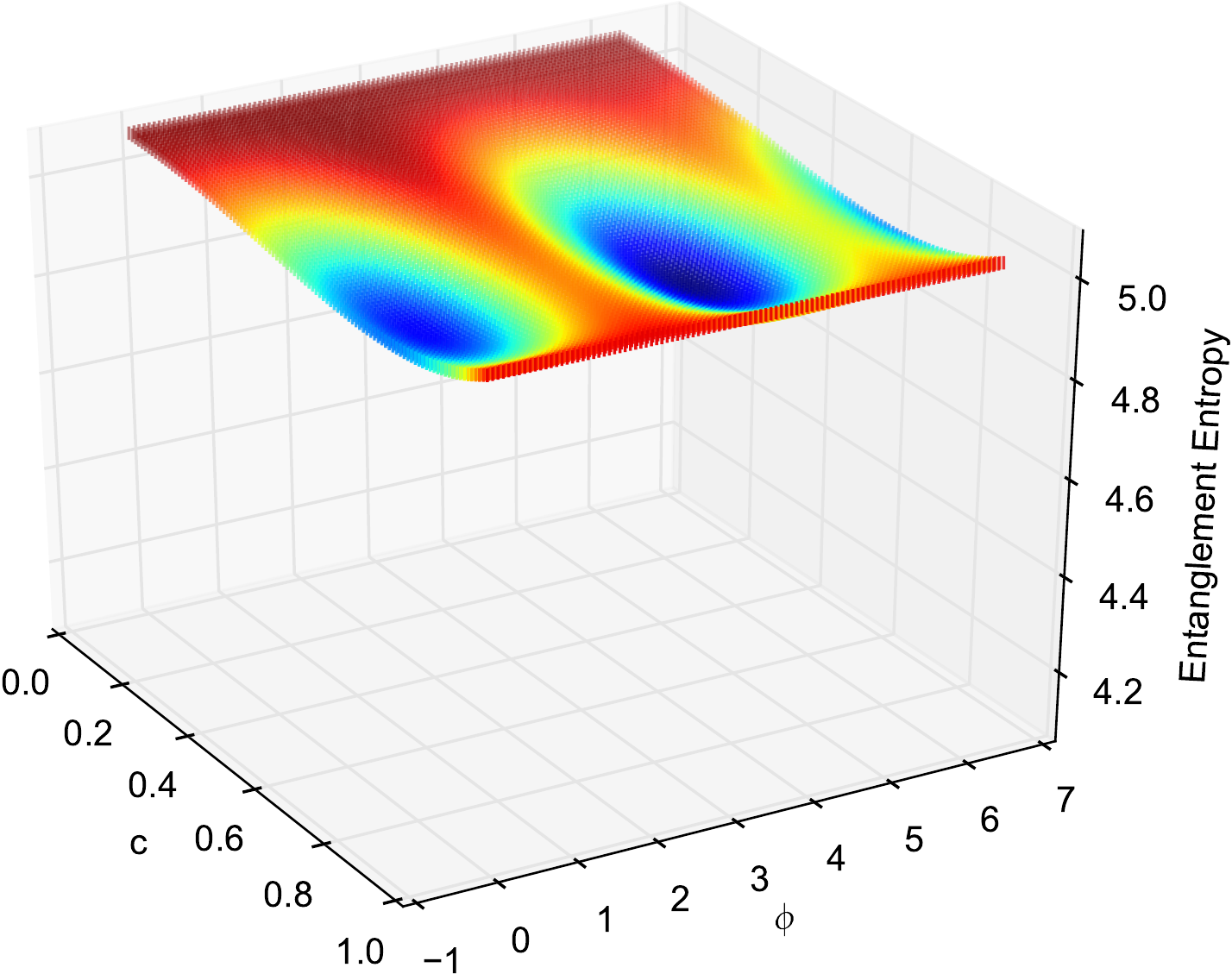}
\includegraphics[width=0.49\linewidth]{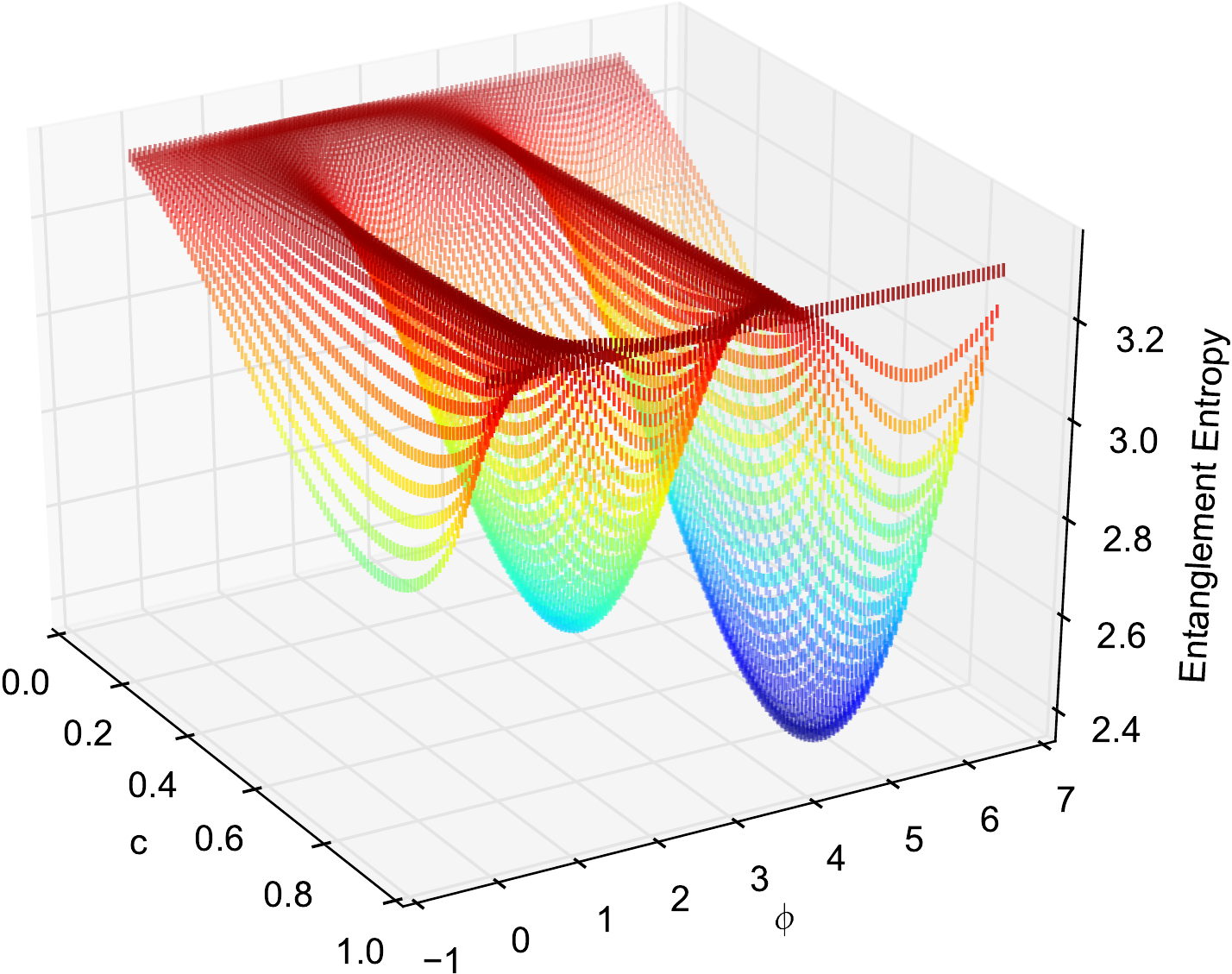}
\includegraphics[width=\linewidth]{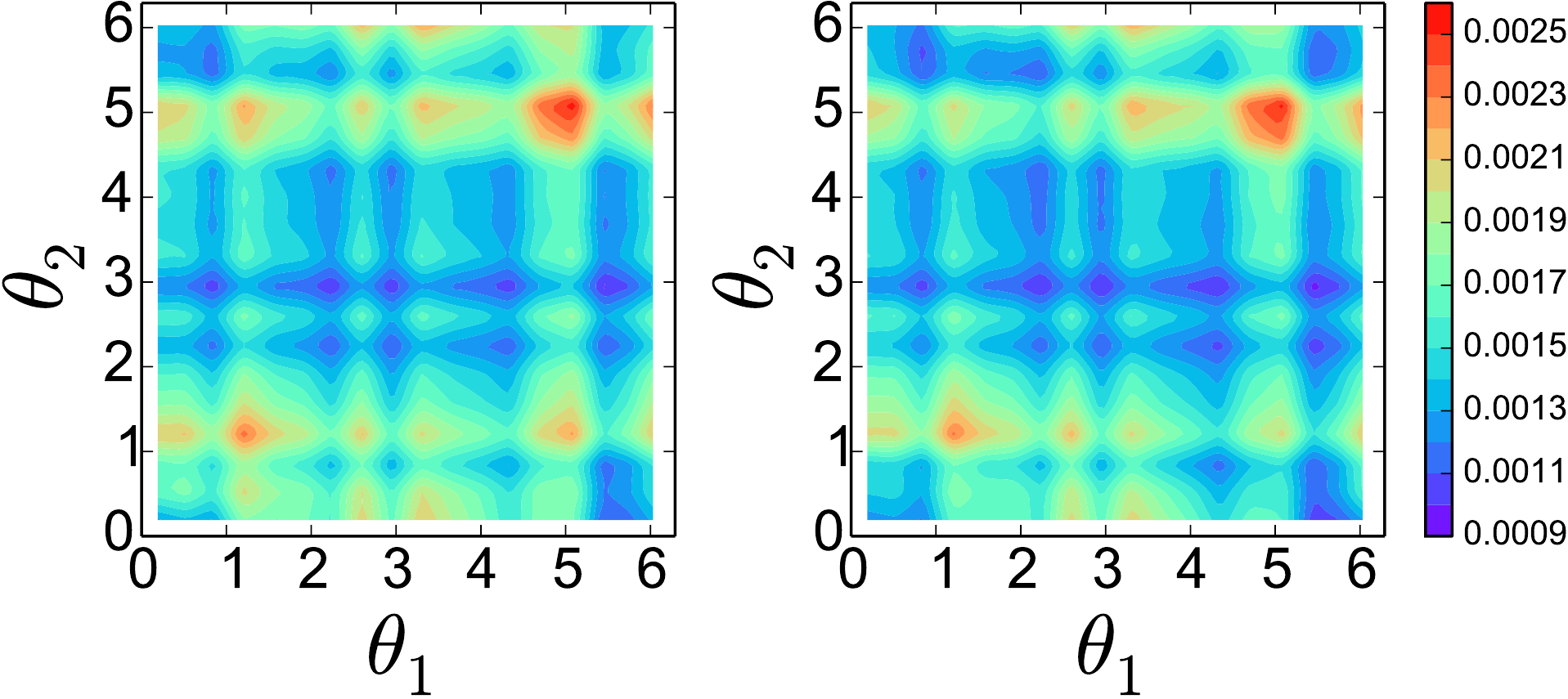}
\caption{Renyi entanglement entropy for the $42$a cluster obtained from the reduced density 
matrix along two topologically non-trivial cuts. Along the first cut (top left), 
the entanglement entropy is found not to vary significantly which renders the evaluation of 
the minimally entangled states along this cut unreliable. 
The location of the minima along the second cut (top right) are in 
agreement with the theoretical expectation. However, the Chern number for the 
topological states in $K=(0,0)$ and $K=(\pi,\pi)$ sums to 1.} 
\label{fig:42a_mes}
\end{figure}

In section~\ref{sec:mes} we discussed the case of the 42a cluster
which naively did not yield orthogonal states. 
On probing the Renyi entanglement entropy as a function of parameters 
specifying the minimally entangled states (MES), shown in Fig.~\ref{fig:42a_mes}, 
we observed that while the local minima of the entanglement landscape along 
one cut is deep, along the other cut the region of local minima 
appears fairly flat with no prominent minima with 
the difference between the maxima and minima of $S_2$ being $\approx 0.1$. 
This makes the identification of the MES and hence the evaluation of the $\mathcal{S}$ 
matrix unreliable. This may be due to the small width of the region 
(topologically non-trivial cut) used to compute the reduced density matrices.

\section{Distribution of Chern number at $J_{\chi} = 0$}
\label{sec:dist_chern}

\begin{figure}
\centering
\includegraphics[width=\linewidth]{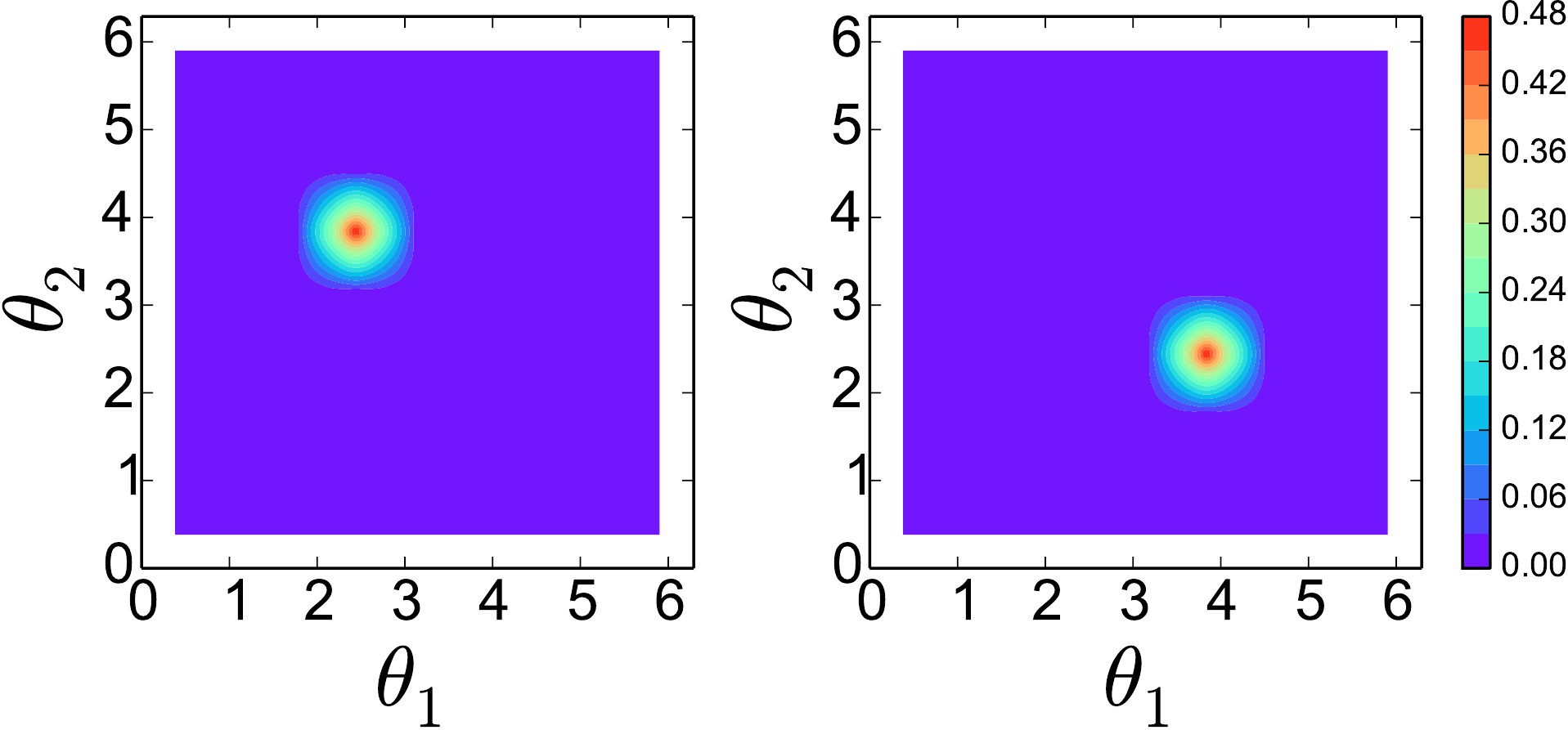}
\includegraphics[width=\linewidth]{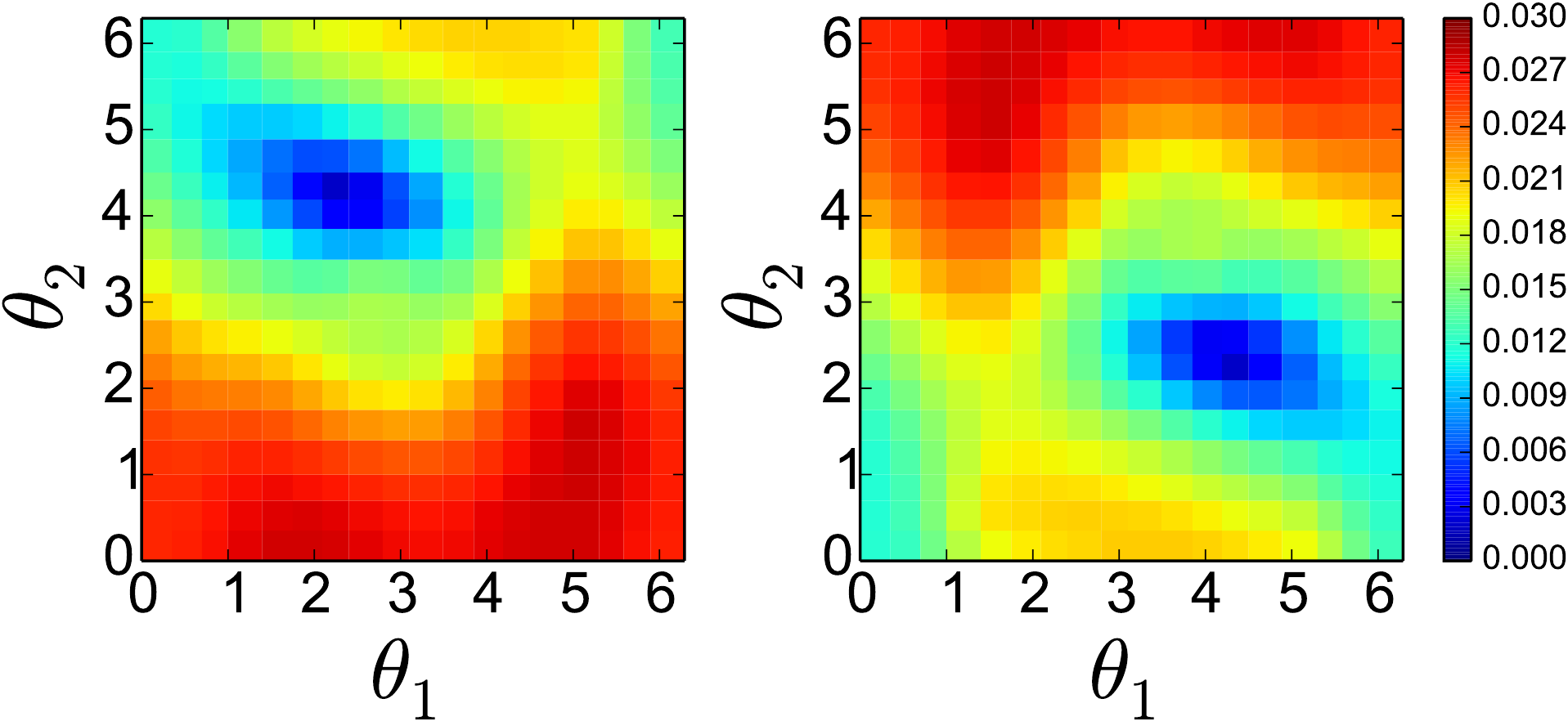}
\caption{Top: Berry curvature as a function of $\theta_1$ and $\theta_2$ for the ground 
states in $K=(0,0)$ (left) and $K=(\pi,0)$ (right) sectors in the case of time reversal 
symmetric Hamiltonian $J_{\chi} = 0 $ on the 42b cluster. In both cases, there is 
essentially one point on the entire grid that contributes to the total Chern number; 
the finite spread seen is an artefact of grid resolution and the interpolation used to present the distribution. 
This region is marked by the bright red spot and its Berry flux is equal to 1/2. 
Below: The energy difference between the first two states in each momentum sector. 
The point of touching of the two energy levels coincides with the location of high Berry 
curvature confirming it to be the source of the non-zero Chern number.}
\label{fig:42b_berry_flux}
\end{figure}

\begin{figure*}
\centering
\includegraphics[width=0.22\linewidth]{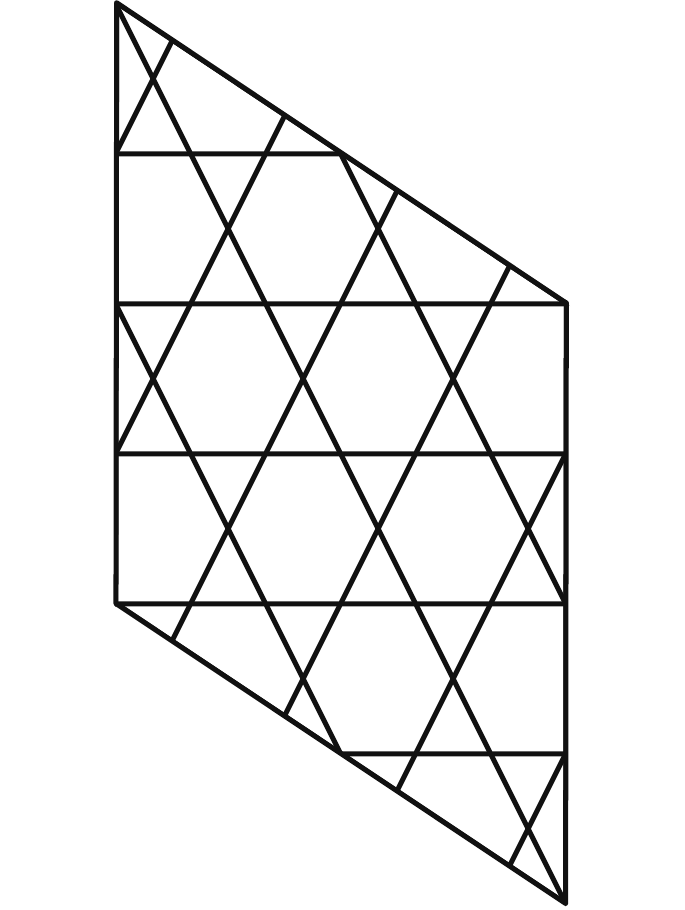}
\includegraphics[width=0.38\linewidth]{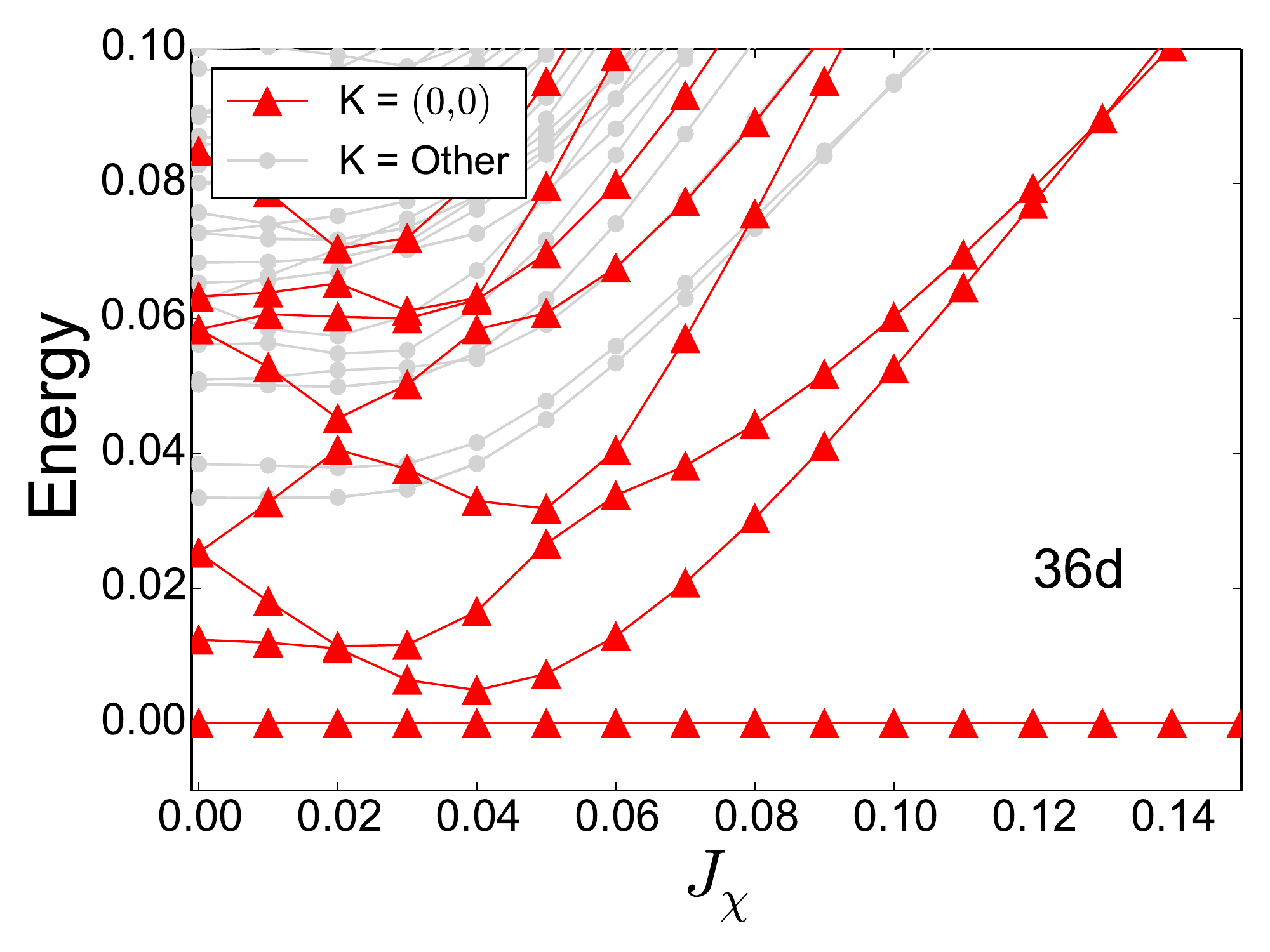}
\includegraphics[width=0.38\linewidth]{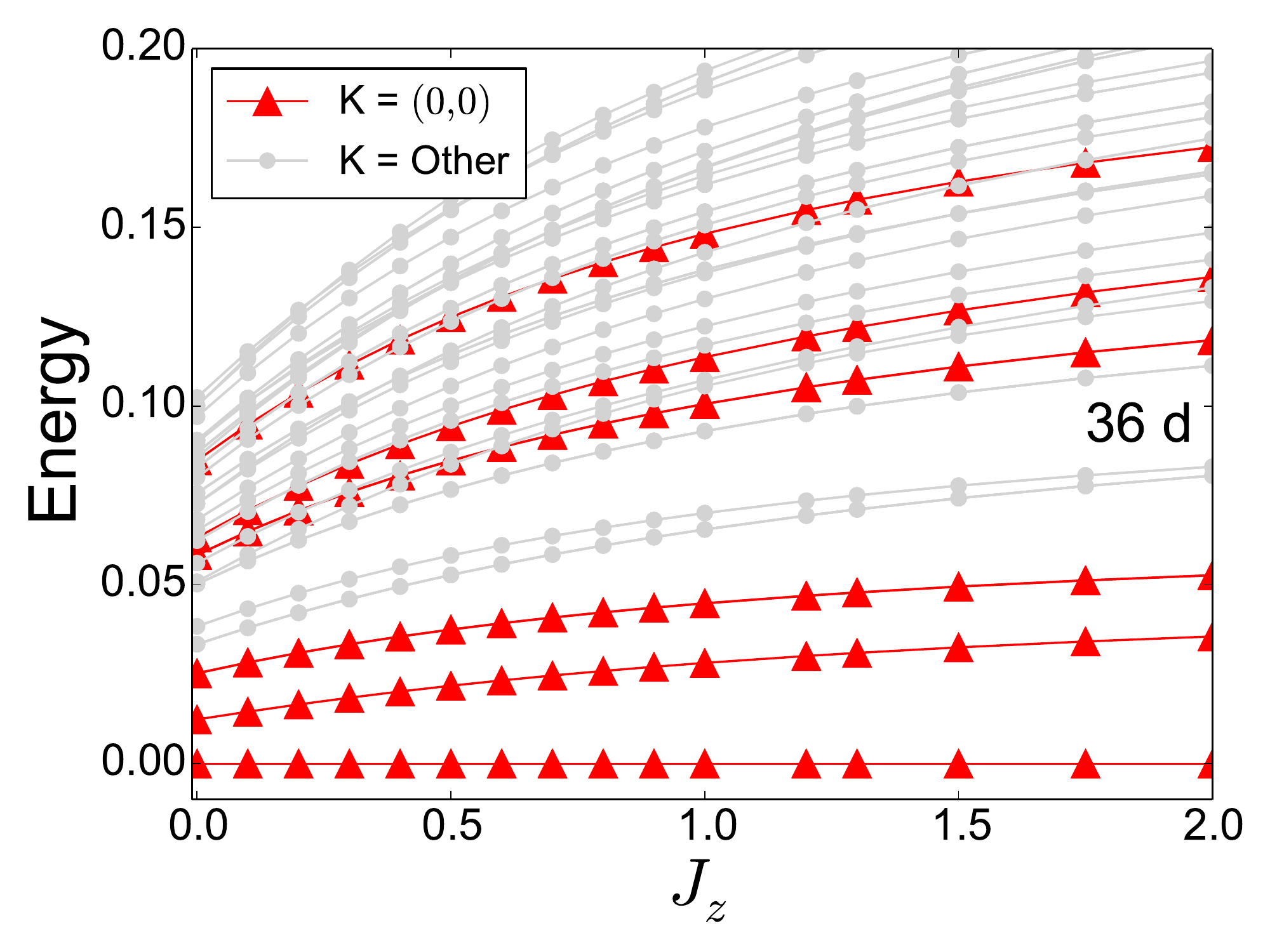}
\caption{Left: geometry of the 36d cluster. The energy spectrum as a function of (center) 
$J_{\chi}$ at $J_z = 0$ and (right) $J_z$ at $J_{\chi} = 0 $ are shown.}
\label{fig:36d_analysis}
\end{figure*} 

Consider a non-interacting band theory, where two bands 
touch at some point in momentum space, and assume this touching is 
protected by time-reversal. When a chiral term (time reversal symmetry breaking term) 
is added the bands do not touch any more, a gap at the 
special point is introduced and each band individually 
acquires a Chern number. Thus, in some sense 
the source of non-zero Chern number can be attributed to be emerging 
from the band touching in a time reversal symmetric model. 

As explained in the text, with zero external chirality, the $XXZ$ Hamiltonian 
is ``time reversal symmetric''. The analogous observation in the case of 
many-body wavefunctions is that the source of non-zero Chern number is the ground and first excited 
state wavefunctions touching at one (or more) points in the two-dimensional space of twist angles, 
one can think of this touching (or crossing) as a state meeting its chiral partner. 
As is seen in Fig.~\ref{fig:42b_berry_flux} for the $42$b cluster,  
we find a Berry monopole (a singular region which has non zero Berry curvature) 
when the lowest state in $K = (0,0)$ is tracked; its location coincides with that of 
the twist at which the energy gap (in that momentum sector) 
goes to zero. 

Note however, the phase assignment of the Berry monopole for the wavefunctions is 
ambiguous ($\pm\pi$) in the absence of time reversal symmetry. 
With the inclusion of the chiral term the time-reversal is broken, the 
Berry monopole spreads out (as is seen in Fig.~\ref{fig:chern_distribution} 
in the main text) and the sign of the Chern number is determined.

\begin{figure*}
\centering
\includegraphics[width=0.32\linewidth]{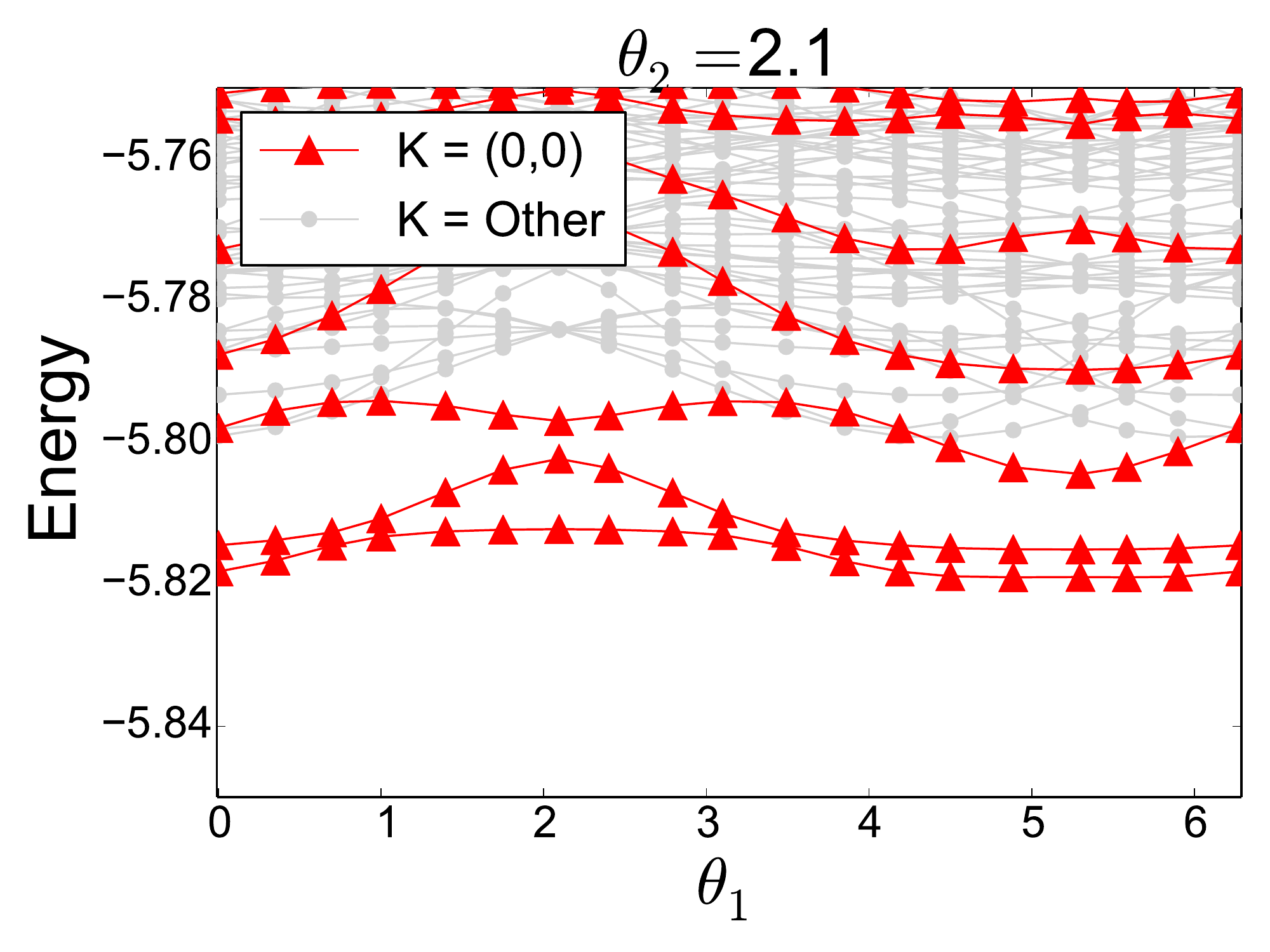}
\includegraphics[width=0.32\linewidth]{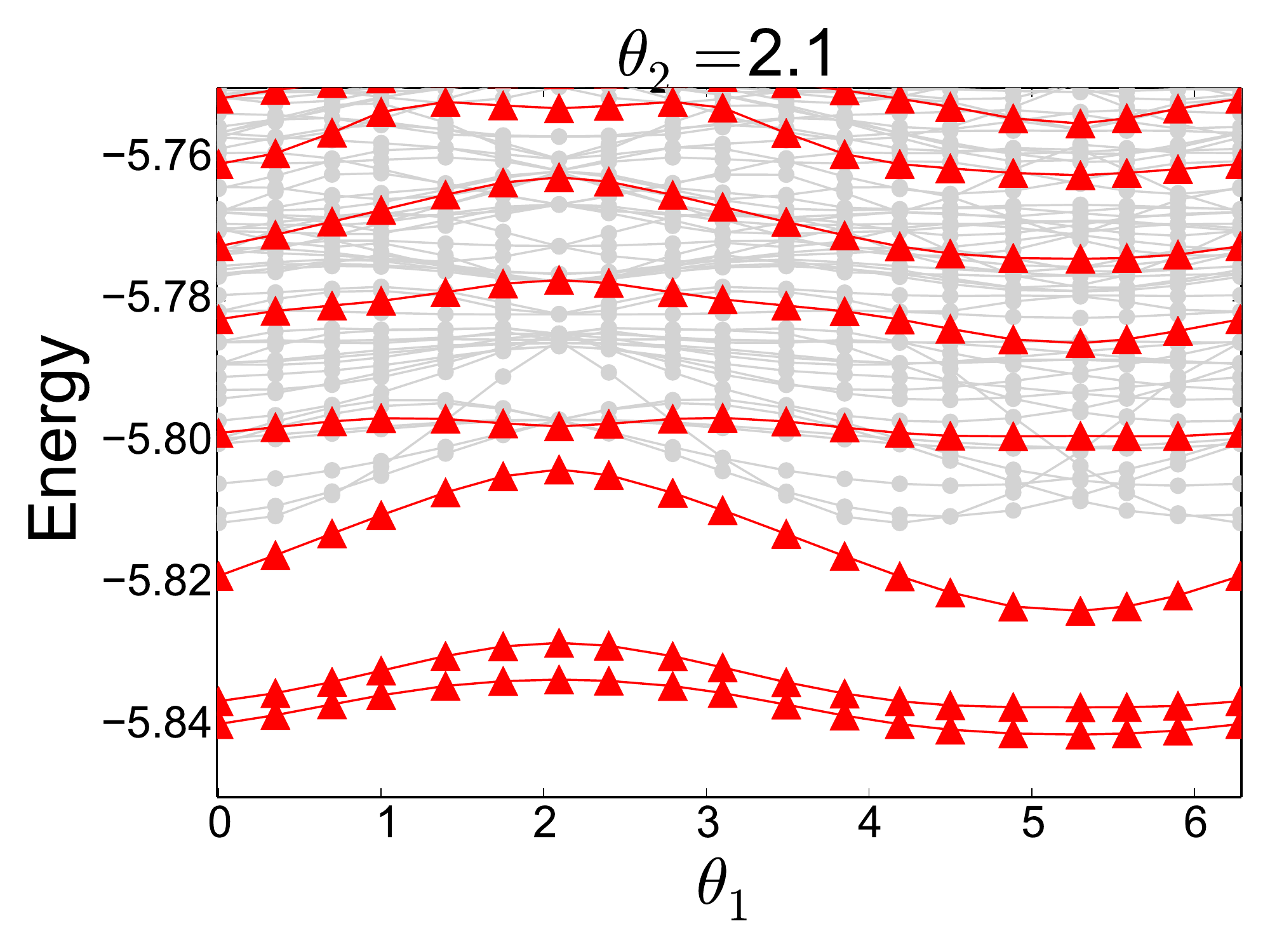}
\includegraphics[width=0.32\linewidth]{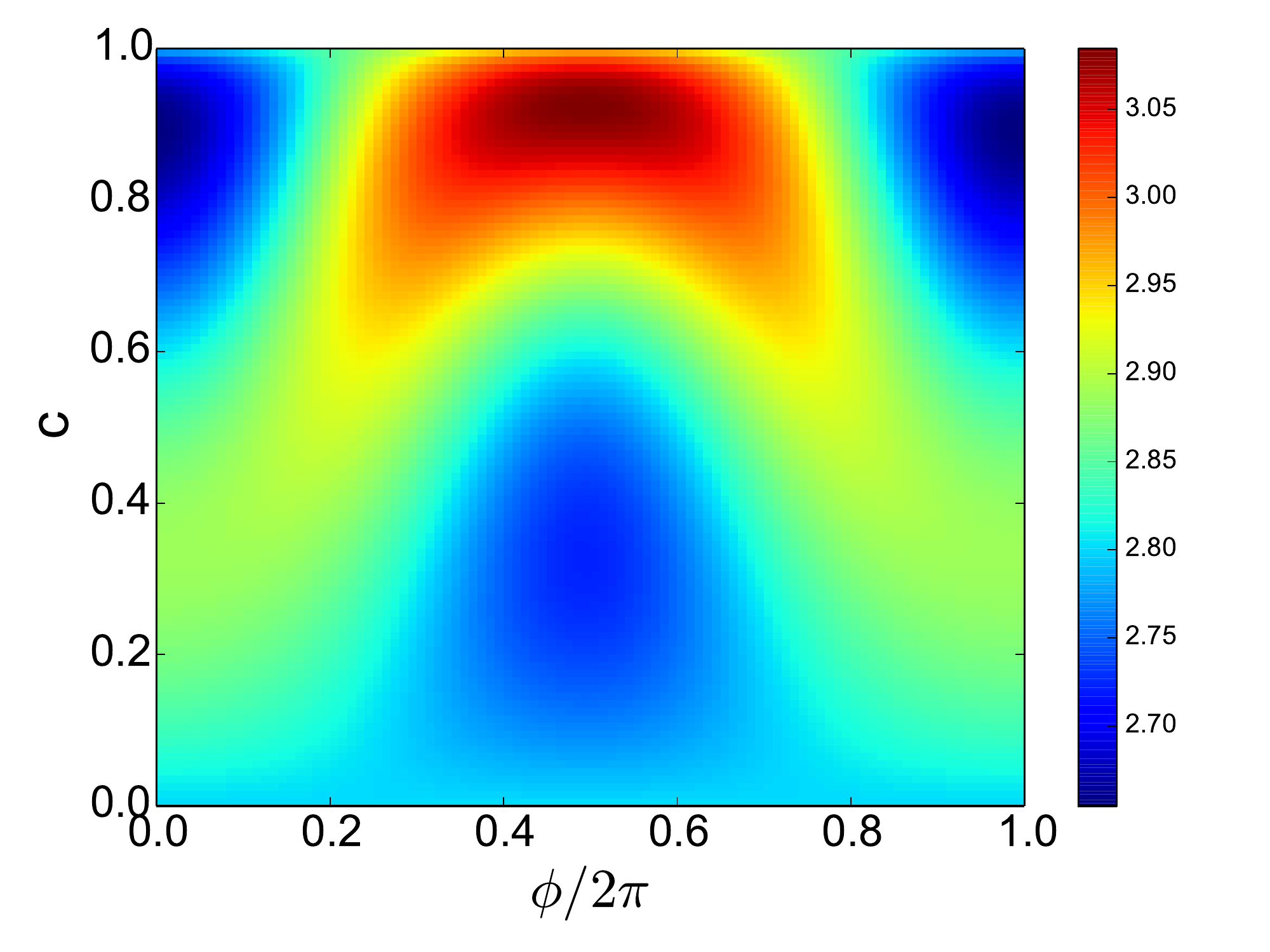}
\caption{Spectral flow of low energy wavefunctions for the 36d cluster 
under twist boundary conditions (varying $\theta_1$) at fixed $\theta_2 = 2.1$ for $J_{\chi}=0.01$ 
(left) and $J_{\chi}=0.04$ (center). The former case shows no clear separation of a 
low energy manifold from the rest of the spectrum, while the latter case shows two topological states. 
On the right, for $J_{\chi}=0.04$, we show the profile of the Renyi entanglement entropy 
for a topologically non trivial cut as a function of parameters $c,\phi$ 
entering the linear combination of the two topological states. 
Two local minima are detected as seen by the blue regions. 
The other topologically non trivial cut (not shown), 
being related by symmetry, has minima at the same location of $c$, 
albeit with different $\phi$.}
\label{fig:36d_flux_mes}
\end{figure*}

\section{Analysis of the 36d cluster}
\label{sec:analysis_36d}  

In this appendix we compile all the results for a single cluster, $36$d, shown in Fig.~\ref{fig:36d_analysis},  
and discuss the results together; we find that this cluster is similar to the 48 cluster.
This cluster has been  
studied by several authors~\cite{Leung1993,Nakano2011,Lauchli2011} primarily in the context 
of the Heisenberg point of the KAF.

Fig.~\ref{fig:36d_analysis} shows the energy spectrum as a function of $J_{\chi}$ 
at $J_z = 0 $. At $J_{\chi}=0$ there are four low energy states (all in $K=(0,0)$) i.e. 
two states followed by an exactly degenerate pair. On turning on $J_{\chi}>0$, 
the degenerate pair splits, one of them approaches the 
ground state  becoming quasi-degenerate with the ground state at $J_{\chi} \approx 0.04$. 
The first excited state at $J_{\chi}$ appears to eventually gap
out around $J_{\chi} \approx 0.04$ and beyond. 

We checked where the ground state manifold is clearly separated out from the continuum
as a function of twisting,
finding (for at least  the part of the manifold we examined)
it is at $J_{\chi}=0.04$ but not at $J_{\chi}=0.01$ (see Fig.~\ref{fig:36d_flux_mes}).
Where the ground state manifold is gapped out, it is sensible to compute the Chern number 
(using the lowest two states within the non-abelian formalism).  We find that $C=1$ for the whole
manifold, to at least six digits, which corresponds to $C=1/2$ per state.

We now address the evaluation of the $\mathcal{S}$ matrix. 
Fig.~\ref{fig:36d_flux_mes}, shows the Renyi entanglement entropy 
of a topologically non trivial cut 
as a function of $c$ and $\phi$, the parameters that characterize 
an arbitrary (normalized) linear combination of two eigenstates.
Two local minima are observed (see Fig.~\ref{fig:36d_flux_mes}) but they
are not orthogonal, the overlap is about 0.147.  
While the minimum at $c\approx 0.89$ is prominent, the one at 
$c \approx 0.32$ is located in a relatively flat basin, analogous to the 
observation in the case of the $42$a cluster. 

\end{appendix}

\bibliographystyle{prsty}
\bibliography{biblio-ED}

\end{document}